\documentclass[preprint,review,12pt]{elsarticle}

\usepackage{amssymb}
\usepackage{amsthm}
\usepackage{amsmath}
\usepackage{multirow}
\usepackage{booktabs}
\usepackage{graphicx}
\usepackage{float}
\usepackage{url}
\usepackage{epstopdf}
\usepackage{pdflscape}
\usepackage{amsthm}
\usepackage{amsmath}
\usepackage[english]{babel}
\usepackage{subfigure}
\usepackage{threeparttable}

\usepackage{algorithm,algorithmicx,algpseudocode}

\usepackage{xcolor}

\usepackage{tabularx}
\usepackage{graphicx}
\usepackage{caption}
\usepackage{threeparttable}
\usepackage[figuresright]{rotating}
\usepackage[toc,page,title,titletoc,header]{appendix} 

\makeatletter

\newcommand{\Rmnum}[1]{\expandafter\@slowromancap\romannumeral #1@}
\makeatother

\journal{XXX}

\begin{document}
\begin{frontmatter}
\title{A Consistent Parallel Isotropic Unstructured Mesh Generation Method based on Multi-phase SPH}

\author[1]{Zhe Ji\corref{mycorrespondingauthor}}
\ead{zhe.ji@tum.de}

\author[2]{Lin Fu\corref{mycorrespondingauthor}}
\ead{linfu@stanford.edu}
\cortext[mycorrespondingauthor]{Zhe Ji and Lin Fu contributed equally to this work}

\author[1]{Xiangyu Hu}
\ead{xiangyu.hu@tum.de}

\author[1]{Nikolaus Adams}
\ead{nikolaus.adams@tum.de}

\address[1]{Chair of Aerodynamics and Fluid Mechanics, Department of Mechanical Engineering, Technical University of Munich, 85748 Garching, Germany}

\address[2]{Center for Turbulence Research, Stanford University, Stanford, CA 94305, USA}

\begin{abstract}

In this paper, we propose a consistent parallel unstructured mesh generator based on a multi-phase SPH method. A set of physics-motivated modeling equations are developed to achieve the targets of domain decomposition, communication volume optimization and high-quality unstructured mesh generation simultaneously. A unified density field is defined as the target function for both partitioning the geometry and distributing the mesh-vertexes. A multi-phase Smoothing Particle Hydrodynamics (SPH) method is employed to solve the governing equations. All the optimization targets are achieved implicitly and consistently by the particle relaxation procedure without constructing triangulation/tetrahedralization explicitly. The target of communication reduction is achieved by introducing a surface tension model between distinct partitioning sub-domains, which are characterized by colored SPH particles. The resulting partitioning diagram features physically localized sub-domains and optimized interface communication. The target of optimizing the mesh quality is achieved by introducing a tailored equation-of-state (EOS) and a smooth isotropic kernel function. The mesh quality near the interface of neighboring sub-domains is improved by gradually removing the surface-tension force once a steady state is achieved. The proposed method is developed basing on a new parallel environment for multi-resolution SPH to exploit both coarse- and fine-grained parallelism. A set of benchmarks are conducted to verify that all the optimization targets are achieved consistently within the current framework.

\end{abstract}


\end{frontmatter}

\section{Introduction}
\label{S:introduction}

The topic of parallel mesh generation is critical for capturing complex physical phenomena in various areas, e.g. Finite Element Analysis (FEA) \cite{frey2007mesh}, Computational Fluid Dynamics (CFD) \cite{park2016unstructured} and image discretization in bioinformatics \cite{feng2015scalable}. Developing scalable, stable and high-quality parallel mesh generation methods is important in reducing simulation cost and achieving high-accuracy for the underlying numerical methods. Recently, new challenges have raised for parallel mesh generation methods due to the rapidly growing capabilities and capacities of modern supercomputers. To fully exploit the potential of distributed memory system, the parallel mesh generator needs to resolve various difficulties, e.g. the data dependency, the load balancing and the irregular behavior of the mesh refinement \cite{feng2016hybrid}. According to the NASA CFD vision 2030 study \cite{slotnick2014cfd}, mesh generation is still a significant bottleneck in CFD and more research is needed.

Sequential unstructured mesh generation methods can be roughly categorized as advancing front methods (AFT) \cite{schoberl1997netgen}\cite{lohner2014recent}, Delaunay refinement-based methods \cite{shewchuk2002delaunay}\cite{chew1997guaranteed}, Delaunay variational-based methods \cite{ni2017sliver}\cite{du1999centroidal}, Particle-based methods \cite{FU2019396}\cite{zhong2013particle}\cite{FU2019396AN}, etc. The advancing front method generates the mesh from the geometry boundary and inserts layers of vertices representing the front iteratively towards the interior of the domain \cite{schoberl1997netgen}. The Delaunay refinement-based method starts from a coarse representation of the geometry and improves the mesh quality by gradually inserting new Steiner points into the domain until a prescribed criterion is achieved \cite{shewchuk2002delaunay}. As for Delaunay variational-based method and particle-based method, either an energy function \cite{ni2017sliver}\cite{du2003tetrahedral} or a target mesh-size function \cite{FU2019396}\cite{persson2006mesh} is defined. Different numerical approaches are then applied to minimize the energy or interpolation error in order to optimize the mesh quality. Despite the similarities shared by both methods, one fundamental difference is whether the connectivity information is required during the optimization procedure. For particle-based mesh generation methods \cite{FU2019396}\cite{zhong2013particle}\cite{meyer2005robust}, pair-wise forces are defined between interacting particles to relax the system towards the target distribution. Therefore, the mesh quality is improved implicitly without constructing a Delaunay tessellation for each optimization iteration. Moreover, since the interaction is constrained locally within a short cutoff radius, only local information is required for each particle. Benefiting from aforementioned unique features, the particle-based method can be easily extended to parallel systems and achieve scalable performance.

Comparing to sequential mesh generator, additional targets arise when the mesh is generated in a distributed memory system. Ideally, a parallel mesh generation method should retain the mesh quality generated by the state-of-the-art sequential code and achieve fully code reuse without significantly deteriorating the scalability of the code \cite{tsolakis2019parallel}. Therefore, in order to accomplish the additional goals in a parallel environment, a consistent formulation is required to maintain the quality of the mesh in a parallel environment.

In the past decades, tremendous efforts have been done to develop parallel unstructured mesh generation methods \cite{chrisochoides2006parallel}. Initially, most of the developed schemes follow a coarse-grained parallel strategy \cite{rakotoarivelo2016fine}. A Domain Decomposition (DD) step is first used to partition the geometry into either continuous sub-domains or discrete simply-connected sub-meshes \cite{lohner1999parallel}\cite{linardakis2004parallel}. Different sequential mesh-generation kernels are applied to mesh each sub-problem and optimize the interface between sub-domains/sub-meshes respectively. The main effort to increase the parallel efficiency relies on the amount of communication required on the sub-problem interfaces \cite{loseille2015parallel}. Depending on the data synchronization strategy, the parallel mesh generations can be categorized into tightly-coupled, partially-coupled and decoupled approaches \cite{chrisochoides2006parallel}. Tightly-coupled methods, e.g. Parallel Optimistic Delaunay Mesh (PODM) \cite{nave2004guaranteed} and Parallel Advancing Front Technique for shared memory computers (PAFT$_{SM}$) \cite{lohner1999parallel}, optimize the mesh in the interior and on the interface of each individual sub-problem simultaneously. While stability and quality of the resulting mesh are guaranteed, the parallel implementation induces a significant amount of communication overhead and features low code reuse. For decoupled approaches, e.g. Parallel Projective Delaunay Meshing (P$^2$DM) method \cite{galtier1996prepartitioning} and Parallel Delaunay Domain Decoupling (PD$3$) method \cite{linardakis2004parallel}, the geometry first is decomposed into continuous sub-domains and each sub-domain is meshed separately. This approach achieves high code reuse, but the mesh quality depends on a proper domain decomposition method. The partially coupled strategy, e.g. Parallel Octree AFT (POAFT) method \cite{lohner1999parallel} and Parallel Constrained Delaunay Meshing method (PCDM) \cite{chew1997parallel}, find a balance between the aforementioned two approaches. The meshing procedure is separated into two phases by defining an interior region and interface region. The amount of communication is significantly reduced comparing to tightly coupled ones and the codes are more stable in terms of achieving high-quality meshes.

The coarse-grained mesh generators generally feature irregular communication patterns and lack data locality due to the excessive remote data access \cite{feng2016two}. Recently with the fast development of manycore processors, e.g. Graphic Processing Unit (GPU), the coarse-grained schemes are no longer suitable for the newly-emerged architectures \cite{remacle2015two}. Several fine-grained parallel models are exploited to achieve higher concurrency and data locality in shared-memory systems \cite{rokos2015thread}\cite{rakotoarivelo2016fine}. In \cite{rakotoarivelo2016fine}, the data dependency and concurrency are ensured by constructing a graph and utilizing a fine-grained edge-coloring algorithm respectively. Apart from fine-grained parallel models, a hybrid two-level Locality-Aware Parallel Delaunay imaging-to-mesh conversion algorithm (LAPD) is developed in \cite{feng2016two}. A partially coupled scheme is employed operating at the coarse level, and a tightly coupled method PODM is utilized to optimize mesh quality within each sub-domain. The inter-node communication only happens at the coarse level and high-concurrency is maintained by the tightly coupled approach. More recently, a nested master-worker communication model is proposed in \cite{feng2018hybrid} to overlap the communication and computation and to further exploit the two-level parallelism on manycore distributed memory system.

To conclude, in order to utilize the full potential of the state-of-the-art clusters, the parallel mesh generation method should be able to achieve the following characteristics: (1) well-balanced load, optimized communication volume, high scalability in the node level; (2) high concurrency and data locality property in the thread level within each node; (3) easy to be extended in a parallel environment, i.e. high code reusability.

According to the above discussions, the particle-based mesh generation method is suitable as a candidate of large-scale parallel-mesh generator. Since the mesh generation procedure is accomplished implicitly without operating on a mesh and the pair-wise interaction only relies on its local information, it fulfills the fine-grained parallelism naturally. With a proper domain decomposition method and dynamic load balancing strategy, e.g. the Centroidal Voronoi Particle (CVP) method \cite{fu2017physics}, scalable performance can be achieved with a large number of computational nodes. Moreover, due to the Lagrangian nature of the particle-based method it is particularly suitable and easy to program for modern parallel environment consisting of shared-memory or distributed-memory systems utilizing various parallel techniques, e.g. Message Passing Interface (MPI) \cite{MPI}, OpenMP \cite{openmp08} and CUDA \cite{Nickolls:2008:SPP:1365490.1365500}. A number of well-established codes has been developed for different particle methods on different architectures, e.g DualSPHysics \cite{crespo2015dualsphysics} for free-surface weakly-compressible flows using Smoothed Particle Hydrodynamics (SPH), LAMMPS \cite{plimpton1995fast} for Molecular Dynamics (MD) and Dissipative Particle Dynamics (DPD) simulations, OpenFPM \cite{incardona2019openfpm} for hybrid particle-mesh simulations, etc. To the best of our knowledge, the topic of parallel particle-based mesh generator in a distributed memory system has not yet been explored.

In this paper, a consistent particle-based parallel unstructured mesh generation method is developed. Unlike other parallel approaches, which rely on a domain decomposition step first before generating the mesh, the proposed method merges both steps into a single phase. The targets of improving mesh quality, optimizing communication volume and domain decomposition are achieved consistently within one set of physics-motivated modeling equations. By defining a unified target function and introducing a surface-tension model in the governing equation, the newly-developed SPH-based isotropic unstructured mesh generation method \cite{FU2019396} is extended to a parallel multi-resolution environment \cite{ji2018new}\cite{linfu_fast_neighbor}. The parallel framework employs both MPI and Thread Building Blocks (TBB) \cite{contreras2008characterizing} techniques, therefore the mesh-generation procedure is able to exploit the parallelism with both coarse- and fine-grained abstractions.

The rest of the paper is arranged as follows: In Section \ref{S:model_equations}, we first introduce the mathematical description of our targets. The main idea and the physics-motivated modeling equations are then elaborated. The detailed numerical methods, e.g. the geometry definition, the discretization of the modeling equations, the parallel environment, and etc., are presented in Section \ref{S:numerical_algorithm}. In Section \ref{S:validation}, various validation tests are carried out to demonstrate the performance of the proposed method.

\section{Physics motivated models}
\label{S:model_equations}

\subsection{Target definition}
\label{S:model_equations_Target definition}

We first introduce the mathematical definition of the targets in both the domain decomposition and the mesh generation. Given a target function $\Phi(\textbf{x})$ defined in domain $\Omega$, a point set $V$ is initialized to partition the domain into elements. Each element in the resulting tessellation can be treated as a computational unit. We can characterize the partition as a graph $G=(V,E)$, where $E$ denotes the communication relationship between computational units.

In parallel simulation, $V$ is divided into $n$ disjoint subsets denoted as $V_1$, $V_2$,..., $V_n$ respectively, and each subset is associated with one MPI task. An optimal partitioning should have the following properties \cite{fu2017novel}:
\begin{description}
	\item[$\bullet$] $V_1\cup V_2 ... \cup V_n = V$ and $V_i \cap V_j = \emptyset$ with $i\neq j$;
	\item[$\bullet$] $|V_i|\approx\lceil\dfrac{|V|}{n}\rceil$, $i=1,2,...,n$;
	\item[$\bullet$] $\sum_{i<j}E_{ij}$ is minimum, where $E_{ij}=\textbf{\{}\{u,v\}\in E|u\in V_i \wedge v\in V_j \textbf{\}}$.
\end{description}
Note that in this paper only equal mass partitioning is considered.

Regarding to mesh generation, there exists several approaches to characterize the approximation error between the discretized mesh element and the given target function. According to \cite{ni2017sliver}, the $L^p$ norm between the gradient of $\Phi(\textbf{x})$, $\bigtriangledown\tilde{\Phi}(\textbf{x})$, and its interpolation is defined to characterize the error since the mesh quality can strongly affect the gradient error. In this paper, we define an optimal mesh quality by
\begin{description}
	 \item[$\bullet$] $L(\textbf{x})=\int_{\Omega}\|\bigtriangledown\tilde{\Phi}(\textbf{x})-\bigtriangledown\Phi(\textbf{x})\|_{L^p}d\textbf{x}$ is minimum.
\end{description}

\subsection{Main concept}
\label{S:model_equations_main_idea}

According to the definition in Section \ref{S:model_equations_Target definition}, we propose that the target function $\Phi(\textbf{x})$ can be used for both the domain decomposition and mesh generation. The total number of mesh vertices can be calculated basing on $\Phi(\textbf{x})$, considering that the total mass is known and that each particle has unit mass. The target mass of each sub-domain for load balancing can be determined straightforwardly. A color function can be defined for each particle, where within the same sub-domain particles share the same color.

To achieve the objective of domain decomposition and communication optimization, a surface tension force can be applied between particles with distinct colors to preserve the sharp interface condition between neighboring sub-domains \cite{fu2017novel}. Consequently, particles of the same color tend to concentrate. Sub-domains are optimized towards convex and compact shape due to the existence of surface tension force. According to \cite{fu2017novel}, the steady-state multi-phase fluid has high similarity to the balanced partitioning diagram, and the objectives of domain decomposition and communication can be achieved implicitly.

During the partitioning stage, the mesh quality can be optimized simultaneously. By introducing a tailored equation-of-state (EOS), the relative error of particle density and target density is characterized as pseudo pressure. The error gradient results in pair-wise particle interaction force and drives particles towards target density distribution while maintaining a regularized and isotropic distribution \cite{FU2019396}. Once a steady state has been achieved, the particles in the inner region of each sub-domain are optimized, and the target of minimizing $L(\textbf{x})$ is achieved implicitly.

Last, the mesh quality near the interface region of neighboring sub-domains can be optimized by gradually alleviating the surface tension force. Since an equilibrium state already has been achieved, the optimization of mesh quality near interface region will only result in local redistribution of mesh vertices, i.e. limited overhead of the communication volume.


\subsection{Target function}
\label{S:model_equations_target_function}

We first refer the target function as "target density" function or "density" function to relate with fluid dynamics
\begin{equation}
\label{rho_target_function}
	\rho_t=\Phi(\textbf{x})
\end{equation}
Since the target density function determines the size of mesh elements, we can further define the target feature-size function ($h_t$) based on $\rho_t$ through a mapping function,
\begin{equation}
\label{feature_size_density_equation}
	h_t=Q(\rho_t).
\end{equation}
The target density function can be defined considering different characteristic fields. In general we can write
\begin{equation}
\label{eq:target_mesh_gen}
	\left\{
		\begin{array}{cr}
			\rho_t=\Phi_1(a_1,a_2,...,a_n), & \\
			h_t   =\Phi_2(a_1,a_2,...,a_n), & 
		\end{array}
	\right.
\end{equation}
where $a_1,a_2,...,a_n$ are the contributing factors that characterize the mesh-vertex distribution. In \cite{FU2019396}, the authors suggest to calculate the target function considering the level-set function $\phi$, the diffused curvature $\kappa$ and the minimum distance function $\psi$ taking into account the effect of geometry singularities. Moreover, the target function can also be an arbitrary user-defined function to facilitate capturing details with various objectives.

Based on the target density function, the total mass for generating a volume mesh can be calculated by \cite{FU2019396AN}
\begin{equation}
\label{eq:total_mass_2}
	M_{v}=\int_{\Omega}\rho_t\bigtriangleup dv,
\end{equation}
where
\begin{equation}
\label{eq:triangle_define}
	\bigtriangleup=\left\{
		\begin{array}{cl}
			1, & $if inside the geometry,$\\
			0, & $otherwise.$ 
		\end{array}
	\right.
\end{equation}
The total mass for generating a surface mesh can be calculated similarly by integrating the target density function over the geometry surface
\begin{equation}
\label{eq:total_mass_3}
	M_{s}=\int_{\partial\Omega}\rho_tds.
\end{equation}

In order to characterize the target information for domain decomposition, we define the total computational load ($M_{t}$), i.e. total mass, as
\begin{equation}
\label{eq:total_mass_1}
	M_{t}=M_{v}+M_{s}.
\end{equation}
Then the target mass for each sub-domain can be derived by
\begin{equation}
\label{eq:mass_for_partition}
	M_{proc_i}=\omega_{i,t}M_{t}, 
\end{equation}
where $i=\{1,2,..,N_{proc}\}$. $\omega_{i,t}$ is the fraction of the target mass for sub-domain $i$ compared with the total mass, and $\sum_i^{N_{proc}}\omega_{i,t}=1$. For equal mass partitioning, $\omega_{i,t} = \dfrac{M_{t}}{N_{proc}}$.

\subsection{Model equations}
\label{S:model_equations_model_equations}

The Lagrangian form of governing equations for isothermal multi-phase compressible flows is \cite{fu2017novel}
\begin{equation}
\label{eq:continuity_equation}
	\dfrac{d\rho}{dt}=-\rho\bigtriangledown\cdot\textbf{v},
\end{equation}
\begin{equation}
\label{eq:momentum_equation}
	\dfrac{d\textbf{v}}{dt}=-\textbf{F}_p+\textbf{F}_v+\textbf{F}_s,
\end{equation}
\begin{equation}
\label{eq:moving_equation}
	\dfrac{d\textbf{x}}{dt}=\textbf{v},
\end{equation}
where $\rho$ is the density, $\textbf{v}$ the velocity vector, $\textbf{x}$ the position. $\textbf{F}_p$, $\textbf{F}_v$ and $\textbf{F}_s$ denote the pressure force, the viscous force and the surface tension force respectively.

To close the system, an equation of state (EOS) is required:
\begin{equation}
\label{eq:EOS}
	p=f(\rho),
\end{equation}
where $p$ denotes the fluid pressure. In the current paper, since the particles are treated as pseudo fluid and the objective is to minimize the interpolation error, the equation of state can be set as
\begin{equation}
\label{eq:EOS_current}
	p=P_0\Big(\dfrac{\rho}{\rho_t}\Big)^\gamma,
\end{equation}
where $P_0$ is a constant pressure and $\gamma$ is a user-defined parameter. This EOS drives particles to relax to the target distribution. Once an equilibrium state has been reached, pressure becomes constant, and consequently the interpolation error is minimized, i.e. $\dfrac{\rho}{\rho_t}$ is approximately constant.

The model equations can be discretized and solved by the Smoothed Particle Hydrodynamics method. The discretized form is presented in Section \ref{S:numerical_algorithm_sph_discretization}.

\section{Numerical algorithms}
\label{S:numerical_algorithm}

The main numerical algorithms and implementation details are elaborated in this section.

\subsection{Level-set method for geometry definition}
\label{S:numerical_algorithm_levelset_function}

First, we use the level-set method \cite{osher1988fronts} to represent the geometry surface using a zero level-set following \cite{FU2019396}.
\begin{equation}
\label{eq:level_set_function}
	\Gamma=\{(x,y)|\phi(x,y,t)=0\}.
\end{equation}
The level-set field is discretized on a Cartesian background mesh. The mesh-generation region is defined as the positive phase, i.e. $\Gamma_{+}=\{(x,y)|\phi(x,y,t)>0\}$. 

\subsection{Target information calculation}
\label{S:numerical_algorithm_target_information}

\begin{figure}[H]
  \centering
    \includegraphics[width=0.8\textwidth]{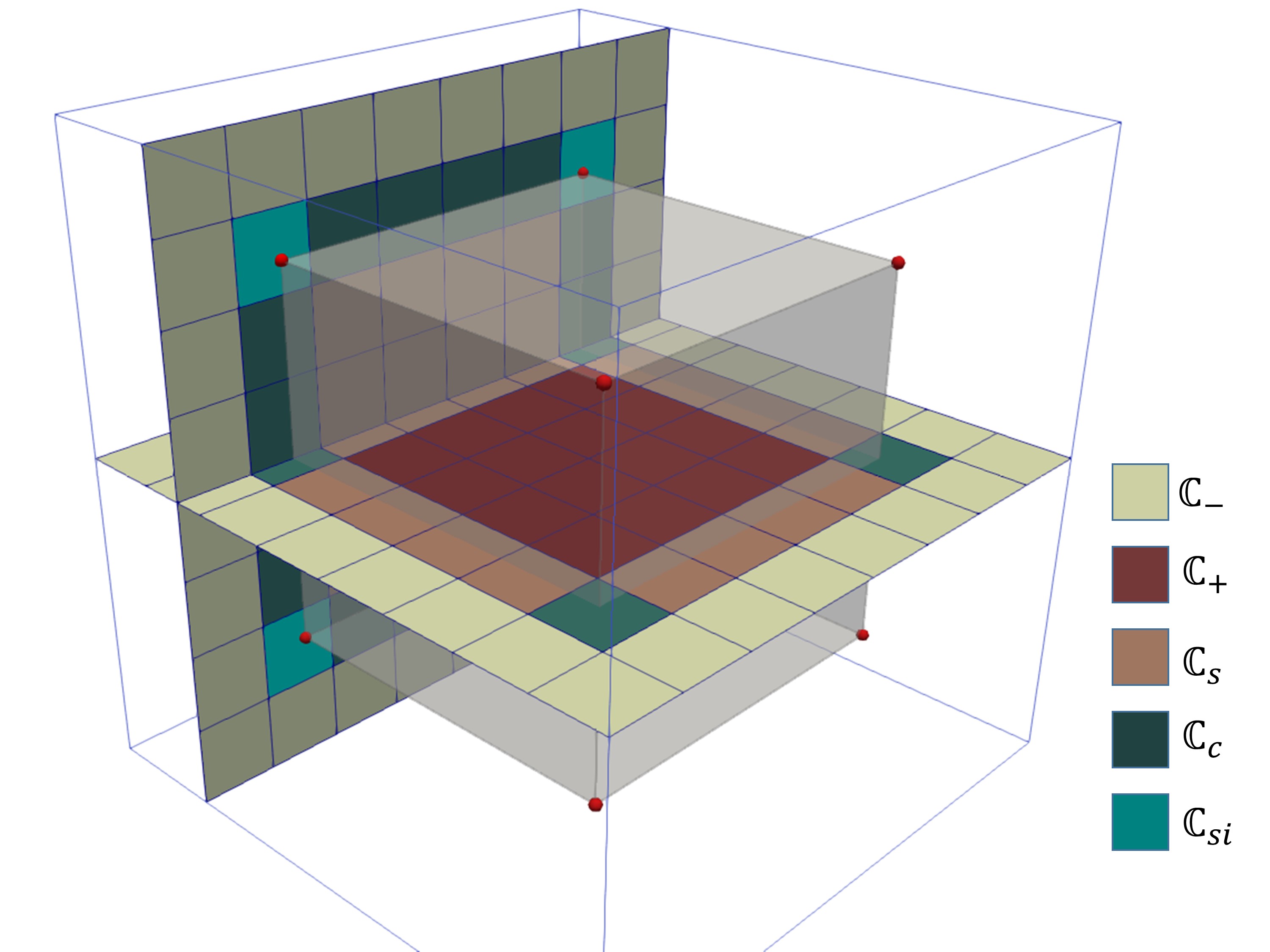}
  \caption{Tag system developed for characterizing the geometry and facilitating the mesh generating process.}
\label{fig:tag_systems}
\end{figure}

The target information defined in Section \ref{S:model_equations_Target definition} is calculated utilizing the same Cartesian background mesh. A tag system is defined to characterize the positive/negative phase, feature curve, feature surface and singularity respectively. Each cell $\mathbb{C}_{i}$ is assigned with a unique integer and five categories are defined accordingly, i.e. positive cell ($\mathbb{C}_{+}$), negative cell ($\mathbb{C}_{-}$), feature-surface cell ($\mathbb{C}_{s}$), feature-curve cell ($\mathbb{C}_{c}$) and singularity cell ($\mathbb{C}_{si}$) (see Fig. \ref{fig:tag_systems}). Moreover, according to the level-set function, the volume fraction of the positive phase inside each cell can be determined explicitly. 

To calculate the target information, the integration can be performed efficiently by looking for corresponding cells based on the tag system. In this paper, the total mass for volume mesh $M_v$ is calculated using the divide-and-conquer method \cite{han2014adaptive} for $\mathbb{S}_v=\{\mathbb{C}_i|\mathbb{C}_i\in(\mathbb{C}_{+}\cup\mathbb{C}_{s}\cup\mathbb{C}_{c}\cup\mathbb{C}_{si})\}$. The total mass for surface mesh $M_s$ is calculated separately considering the surface integral on the feature surface and the line integration on the feature curve. The surface integration is calculated for $\mathbb{S}_s=\{\mathbb{C}_i|\mathbb{C}_i\in(\mathbb{C}_{s}\cup\mathbb{C}_{c}\cup\mathbb{C}_{si})\}$ and the line integration $\mathbb{S}_c=\{\mathbb{C}_i|\mathbb{C}_i\in(\mathbb{C}_{c}\cup\mathbb{C}_{si})\}$. The same divide-and-conquer method can be applied for the integration. To describe the feature surface explicitly, the Marching Cube method \cite{lorensen1987marching} is used for surface reconstruction. Once the total mass for mesh generation $M_t$ is determined, the target mass for partitioning can be calculated straightforwardly applying Eq. (\ref{eq:mass_for_partition}).

In order to determine the total number of SPH particles, we assume that each particle possesses unit mass 
\begin{equation}
\label{eq:mass_SPH_particle}
	m_i=1.
\end{equation}
Since the SPH method inherently conserves mass and there is no mass transfer between particles, the density of each particle will evolve during the relaxation procedure and eventually conforms to the target density field. The target density $\rho_{t,i}$ for particle $i$ is interpolated from the Cartesian background mesh at particle position $\textbf{r}_i$. The target feature-size is calculated following
\begin{equation}
\label{eq:scale_SPH_particle}
	h_{t,i}=\tau\Big(\dfrac{m_i}{\rho_{t,i}}\Big)^{1/d},
\end{equation}
where $\tau$ is a scaling factor depending on the kernel function and $d$ denotes the spacial dimensionality.

\subsection{SPH discretization}
\label{S:numerical_algorithm_sph_discretization}

With standard SPH method, the density of a particle can be calculated using direct summation over all the neighboring $j$ particles 
\begin{equation}
\label{eq:density_sum}
	\rho_i=\sum_jm_jW(r_{ij},h_i),
\end{equation}
where $W(r_{ij},h_i)$ is the kernel function, $h_i$ the smoothing length of particle $i$, $\textbf{r}_{ij}=\textbf{r}_i-\textbf{r}_j$ the connecting vector between particle $i$ and $j$.

Following \cite{FU2019396}, the pressure force can be discretized in a symmetric form as
\begin{equation}
\label{eq:sph_pressure_force_2}
	\textbf{F}_{p,i}=\sum_jm_j\Big(\dfrac{p_0}{\rho^2_{t,i}}+\dfrac{p_0}{\rho^2_{t,j}}\Big)\dfrac{\partial W(r_{ij},h_{ij})}{\partial r_{ij}}\textbf{e}_{ij},
\end{equation}
where $h_{ij}=\dfrac{{h_i+h_j}}{2}$, $\dfrac{\partial W(r_{ij},h_{ij})}{\partial r_{ij}}$ is the derivative of kernel function, and $\textbf{e}_{ij}=\frac{\textbf{r}_{ij}}{r_{ij}}$ is the unit vector pointing from particle $i$ to $j$. 

Note that the density term in the original discretized form disappears by assuming $\gamma=2$ in Eq. (\ref{eq:EOS_current}) (see \cite{FU2019396}), i.e.
\begin{equation}
\label{eq:EOS_2}
	p=p_0\dfrac{\rho^2}{\rho_t^2}.
\end{equation}
Therefore the density summation term defined in Eq. (\ref{eq:density_sum}) is no longer necessary.

The viscous force is calculated following
\begin{equation}
\label{eq:sph_viscous_force}
	\textbf{F}_{v,i}=\sum_jm_j\dfrac{2\eta_i\eta_j}{\eta_i+\eta_j}\Big(\dfrac{1}{\rho^2_i}+\dfrac{1}{\rho^2_j}\Big)\dfrac{\partial W(r_{ij},h_{ij})}{\partial r_{ij}}\dfrac{\textbf{v}_{ij}}{r_{ij}},
\end{equation}
where $\textbf{v}_{ij}=\textbf{v}_{i}-\textbf{v}_{j}$, and $\eta=\rho\nu$ is the dynamic viscosity. In this paper, we use
\begin{equation}
\label{eq:viscosity_coefficient}
	\nu\sim0.1r_c|\textbf{v}|,
\end{equation}
where $r_c$ is the cut-off radius of particle interaction range, assuming that the local Reynolds number is on the order of $O(10)$. Moreover, by setting
\begin{equation}
\label{eq:sph_viscous_force_2}
	\rho=\rho_t,
\end{equation}
the viscous force model can be further simplified. Meanwhile, as suggested in \cite{FU2019396}, a simplified friction model is utilized to set an effectively infinite friction coefficient and to damp particle kinetic energy to zero after each time-step.

As discussed in Section \ref{S:introduction} and Section \ref{S:model_equations}, the targets of maintaining compact and physically-connected sub-domains are handled by introducing a surface tension model between particles belonging to distinct sub-domains, i.e. particles carrying different colors. Ideally, the surface tension force can be modeled by the continuum surface force (CSF) method \cite{brackbill1992continuum} or the continuum surface stress (CSS) method \cite{lafaurie1994modelling}, where a finite transitional band is used to characterize the interface. Within the transitional band the surface-tension force is approximated as a continuous force. However, to avoid the direct calculation of curvature a simplified surface tension model is utilized in this paper. Similarly with \cite{fu2017novel}, the acceleration contributed by the interface force can be approximated by an inter-particle repulsive pressure force,
\begin{equation}
\label{eq:sph_st_force}
	\textbf{F}_{s,i}=-\sum_j\beta(t) m_j\Big(\dfrac{p_0}{\rho^2_{t,i}}+\dfrac{p_0}{\rho^2_{t,j}}\Big)\dfrac{\partial W(r_{ij},h_{ij})}{\partial r_{ij}}\textbf{e}_{ij},
\end{equation}
where $\beta(t)$ is a time-dependent coefficient to characterize the strength of surface-tension effect
\begin{equation}
\label{eq:sph_st_beta}
	\beta(t)=\left\{
			\begin{array}{rcl}
				0, & $ if $C_i = C_j,$ $ &  \\  
				\sigma(t), & $ overwise, $&
			\end{array}
		\right.
\end{equation}
where $C_k$ is the color function for particle $k$. Note that the simplified surface-tension model has high similarity with the discretized pressure force formulation, Eq. (\ref{eq:sph_pressure_force_2}). With the coefficient $\sigma(t)>0$, particles of different colors near the interface region of neighboring sub-domains are separated by a repulsive force. Moreover, in high-curvature regions particles are expected to concentrate more, consequently resulting in a larger repulsive force to regularize the interface shape.

To incorporate with the mesh generation procedure, the surface tension force is removed once the target of domain-decomposition is achieved. In order to maintain numerical stability, $\beta(t)$ is reduced gradually. In this paper, we use a linear function to ramp down the strength of surface tension effect between the time interval $[t_0,t_1]$
\begin{equation}
\label{eq:sph_st_sigma}
	\beta(t)=\left\{
			\begin{array}{ccl}
				3 & , & $ if $t \le t_0,$ $   \\  
				3(1-\frac{t_0-t}{t_0-t_1}) & , & $ if $t_0<t \le t_1,$ $   \\  
				0 & , & $ if $t_1<t,$ $  \\  
			\end{array}
		\right.
\end{equation}
where $t_0$ is the time when the initial partitioning is converged, and an initial constant of 3 is set for $\beta$ following the suggestion in \cite{fu2017novel}. $t_1$ can be obtained by adding a fraction of $t_0$, i.e. $t_1=(1+\vartheta)t_0$. In this paper, we set $\vartheta\in[0.1,0.2]$.

The convergence of partitioning is achieved when the particle system is fully relaxed and an equilibrium state is maintained. We measure the topology variation of the system for a certain amount of iterations, e.g. 50. If the topology remains static for the predefined interval, we assume the partitioning procedure terminates and set $t=t_0$. In order to measure the topology variation, the sampling procedure introduced in \cite{fu2017novel} can be carried out.

\subsection{Time integration}
\label{S:numerical_algorithm_time_integration}

Following \cite{fu2017novel} and \cite{FU2019396}, a simplified time-integration scheme is employed as
\begin{equation}
\label{eq:time_int_01}
	\widetilde{\textbf{v}}_{n+\frac{1}{2}}=\textbf{v}_n+\dfrac{1}{2}\textbf{a}_n\Delta t,
\end{equation}
\begin{equation}
\label{eq:time_int_02}
	\textbf{v}_{n+\frac{1}{2}}=\widetilde{\textbf{v}}_{n+\frac{1}{2}}+\dfrac{1}{2}\widetilde{\textbf{a}}_{n+\frac{1}{2}}\Delta t,
\end{equation}
\begin{equation}
\label{eq:time_int_03}
	\textbf{r}_{n+1}=\textbf{r}_n+\textbf{v}_{n+\frac{1}{2}}\Delta t.
\end{equation}
The acceleration $\textbf{a}_n$ is first calculated from the pressure force $\textbf{F}_p$ and the surface tension force $\textbf{F}_s$ and is used to update the mid-point velocity $\widetilde{\textbf{v}}_{n+\frac{1}{2}}$. Then the viscous force $\textbf{F}_v$ is calculated to obtain the mid-point acceleration $\widetilde{\textbf{a}}_{n+\frac{1}{2}}$ utilizing the predicted velocity $\widetilde{\textbf{v}}_{n+\frac{1}{2}}$. At last the particle position is updated by a full timestep according to the modified velocity $\textbf{v}_{n+\frac{1}{2}}$.

The time-step size of the simulation is calculated with respect to the physical model to maintain numerical stability. In this paper, the time-step size is determined by the CFL criterion, the viscous criterion, and the body force criterion respectively \cite{FU2019396},
\begin{equation}
\label{eq:time_step_size}
	\Delta t = min\Big(0.25\sqrt{\frac{r_c}{|\textbf{a}|}},\frac{1}{40}\frac{r_c}{|\textbf{v}|},0.125\frac{r_c^2}{\nu}\Big),
\end{equation}
where the artificial speed of sound is assumed as $c_s\sim40|\textbf{v}|_{max}$.

\subsection{Singularity, feature curve and feature surface}
\label{S:numerical_algorithm_singularity}

Four types of particles, i.e. singularity particle, feature-curve particle, feature-surface particle and normal particle, are defined to characterize the features of underlying geometry specifically. Singularity particles are employed to represent geometrical singularities such as sharp corners. The position is not updated once a particle is marked as singularity particle. Feature-curve particles and feature-surface particles are used to represent sharp edges and surface of the geometry respectively. Particles that are inside the mesh-generation region, i.e. the positive phase, are referred as normal particles.

The feature-curve particles are used to provide repulsive force for surface and normal particles to prevent penetration. During the triangulation/tetrahedralization process, they are also utilized to generate 1D line mesh. For the force calculation, only the same type of particles or singularity particles are considered within the cut-off radius. The pair-wise interaction force is projected to the tangential direction of the curve $\textbf{T}(\textbf{r}_i)$ at position $\textbf{r}_i$
\begin{equation}
\label{eq:force_curv}
	\textbf{F}_{*,i}=(\textbf{F}_{*,i}\cdot\textbf{T}(\textbf{r}_i))\textbf{T}(\textbf{r}_i).
\end{equation}
After updating position, the particles are projected back to the feature curve at the closest point to preserve the geometry.

Similarly, the feature-surface particles are used as the boundary conditions of normal particles and are also used to generate surface mesh. The contribution from normal particles are excluded in the force calculation of the feature-surface particles. The normal component of the interaction force is ignored
\begin{equation}
\label{eq:force_surface}
	\textbf{F}_{*,i}=\textbf{F}_{*,i}-(\textbf{N}(\textbf{r}_i)\cdot\textbf{F}_{*,i})\textbf{N}(\textbf{r}_i),
\end{equation}
where $\textbf{N}(\textbf{r}_i)$ is the unit normal vector on the surface. To constrain the particle motion on the surface, the updated position is projected back onto the surface by
\begin{equation}
\label{eq:displacement_surface}
	\textbf{r}_i=\textbf{r}_i-\phi_i\textbf{N}(\textbf{r}_i).
\end{equation}

\subsection{Triangulation and tetrahedralization}
\label{S:numerical_algorithm_triangulation}

In the proposed method, since the mesh quality is optimized implicitly without connectivity information, the 2D triangulation or 3D tetrahedralization is only performed for post-processing to visualize mesh elements and calculate mesh quality. For 2D triangulation, the mesh is constructed similarly with \cite{FU2019396} utilizing a local Voronoi tessellation. At the sub-domain boundaries, the ghost buffer particles from neighboring processors are utilized to generate the Voronoi diagram. A pair-wise synchronization is then performed to remove duplicated elements. For 3D tetrahedralization, the open-source code TetGen \cite{si2015tetgen} is used. The flip operations included in TetGen (2-3 flip, 3-2 flip and 4-4 flip) are performed to improve the connectivity. 

\subsection{Mesh quality criterion}
\label{S:numerical_algorithm_mesh_quality}

For isotropic triangular meshes, the mesh quality is quantified by $G=2\sqrt{3}\frac{S}{PH}$ and the angle $\theta$, where $S$ is the triangle area, $P$ the half-perimeter and $H$ the length of the longest edge. $\theta_{max}$, $G_{avg}$ and $G_{min}$/$\theta_{min}$ are the maximum, average and minimum values respectively. $\theta_{min}^{\#}$ is the averaged value of the minimum angle in each triangle. $\theta_{<30}$ is the number of triangle that contains angle smaller than $30^{\circ}$. The distribution of angles is provided too to check the regularity.

For isotropic tetrahedral mesh, the mesh quality is quantified by the dihedral angle $\theta$ and radius ratio $\gamma=3\frac{r_{in}}{r_{circ}}$ respectively, where $r_{in}$ is the inradius and $r_{circ}$ the circumradius of a tetrahedron. $\theta_{max}$, $\gamma_{avg}$ and $\gamma_{min}$/$\theta_{min}$ are the maximum, average and minimum values respectively. $\theta_{min}^{\#}$ is the averaged value of the minimum dihedral angle in each tetrahedron. To evaluate the distribution, diagrams of dihedral angle and radius ratio are provided. The number of slivers are measured by counting the number of tetrahedra with different smallest dihedral angles, i.e. $10^{\circ}$, $20^{\circ}$, $30^{\circ}$ and $40^{\circ}$.

\subsection{Parallel environment for multi-resolution SPH}
\label{S:numerical_algorithm_parallel_environment}

The proposed method is implemented in a newly developed parallel environment for multi-resolution SPH \cite{ji2018new}, which is designed for large-scale simulations with arbitrarily adaptive smoothing-length. The code utilizes a localized nested hierarchical data structure and a tailored parallel fast-neighbor-search algorithm for an efficient construction of ghost buffer particles in remote processors. The communication is optimized by a "diffused graph" strategy, which facilitates the reduction of communication frequency. The framework is parallelized with both MPI and TBB. The framework exhibits scalable performance on current state-of-the-art computer clusters for both uniform and non-uniform particle distributions. The weak scaling reveals that the code scales well up to at least 3584 cores. Good efficiency is achieved for strong scaling tests (scales up to 7168 cores) at various scales. For more detailed description of the framework, the readers are referred to \cite{ji2018new}\cite{ji2019lagrangian}\cite{linfu_fast_neighbor}.

\subsection{Overview}
\label{S:numerical_algorithm_overview}

In this section, a detailed flowchart of the proposed method is summarized in Algorithm \ref{alg:ParMesh}.

\begin{algorithm}[H]\footnotesize
\caption{Flowchart of the parallel mesh generation method}
\label{alg:ParMesh}
\begin{algorithmic}[1]
\State Initialize the background Cartesian mesh;
\State Initialize the level-set function (Eq. (\ref{eq:level_set_function})) and target density function (Eq. (\ref{eq:target_mesh_gen})) basing on the background mesh;
\State Calculate the target information for mesh generation (Eq. (\ref{eq:total_mass_1})) and domain decomposition (Eq. (\ref{eq:mass_for_partition}));
\State Initialize the parallel environment; 
\Comment {\textit{e.g. construct local data structure, build communication topology, allocate resources, and etc.}} 
\While{$t<t_{end}$} 
  \State Define particle target density ($\rho_t$), scale ($h_t$, see Eq. (\ref{eq:scale_SPH_particle})) as well as other information at $\textbf{r}_i^n$;
  \State Refresh data structure and communication topology;
  \Comment {\textit{See \cite{ji2018new} for detailed description}}
  \State Construct ghost buffer particles, and find neighboring particles to construct neighbor list;
  \State Reset particle velocity and forces;
  \State Calculate pressure force $\textbf{F}_p$ (Eq. (\ref{eq:sph_pressure_force_2}));
  \State Calculate surface-tension coefficient (Eq. (\ref{eq:sph_st_beta}) and accumulate surface-tension force $\textbf{F}_s$ (Eq. (\ref{eq:sph_st_force}));
  \State Map $\textbf{F}_p$ and $\textbf{F}_s$ for feature-curve and feature-surface particles (Eq. (\ref{eq:force_curv}) and Eq. (\ref{eq:force_surface}));
  \State Set physical time-step size (Eq. (\ref{eq:time_step_size})) and update the mid-point velocity $\widetilde{\textbf{v}}_{n+\frac{1}{2}}$ (Eq. (\ref{eq:time_int_01}));
  \State Accumulate viscous force (Eq. (\ref{eq:sph_viscous_force_2}));
  \State Set physical time-step size (Eq. (\ref{eq:time_step_size})) and update predicted velocity $\textbf{v}_{n+\frac{1}{2}}$ (Eq. (\ref{eq:time_int_02}));
  \State Update particle position (Eq. (\ref{eq:time_int_03}));
  \State Find particles that are close to the geometry features utilizing the tag system (defined in Section \ref{S:numerical_algorithm_target_information}). Map the new singularity/feature-curve/feature-surface particles into corresponding singularity/feature curve/feature surface (Section \ref{S:numerical_algorithm_singularity});
  \If{Do post-processing}
    \State Generate the corresponding mesh and calculate mesh quality; 
  \EndIf
\EndWhile
\end{algorithmic}
\end{algorithm}

\section{Numerical validation}
\label{S:validation}

In this section, a set of two- and three-dimensional test cases are presented to validate the performance of our method. All cases in this section are simulated on the facilities provided by Leibniz-Rechenzentrum (LRZ). For all the test cases below, we define the communication volume as $N_{eg}/N_{et}$, where $N_{eg}$ is the number of elements that contain vertices of different colors and $N_{et}$ is the total number of elements generated.

\subsection{Circle}
\label{S:validation_circle}

\begin{figure}[H]
  \centering
    \includegraphics[width=0.8\textwidth]{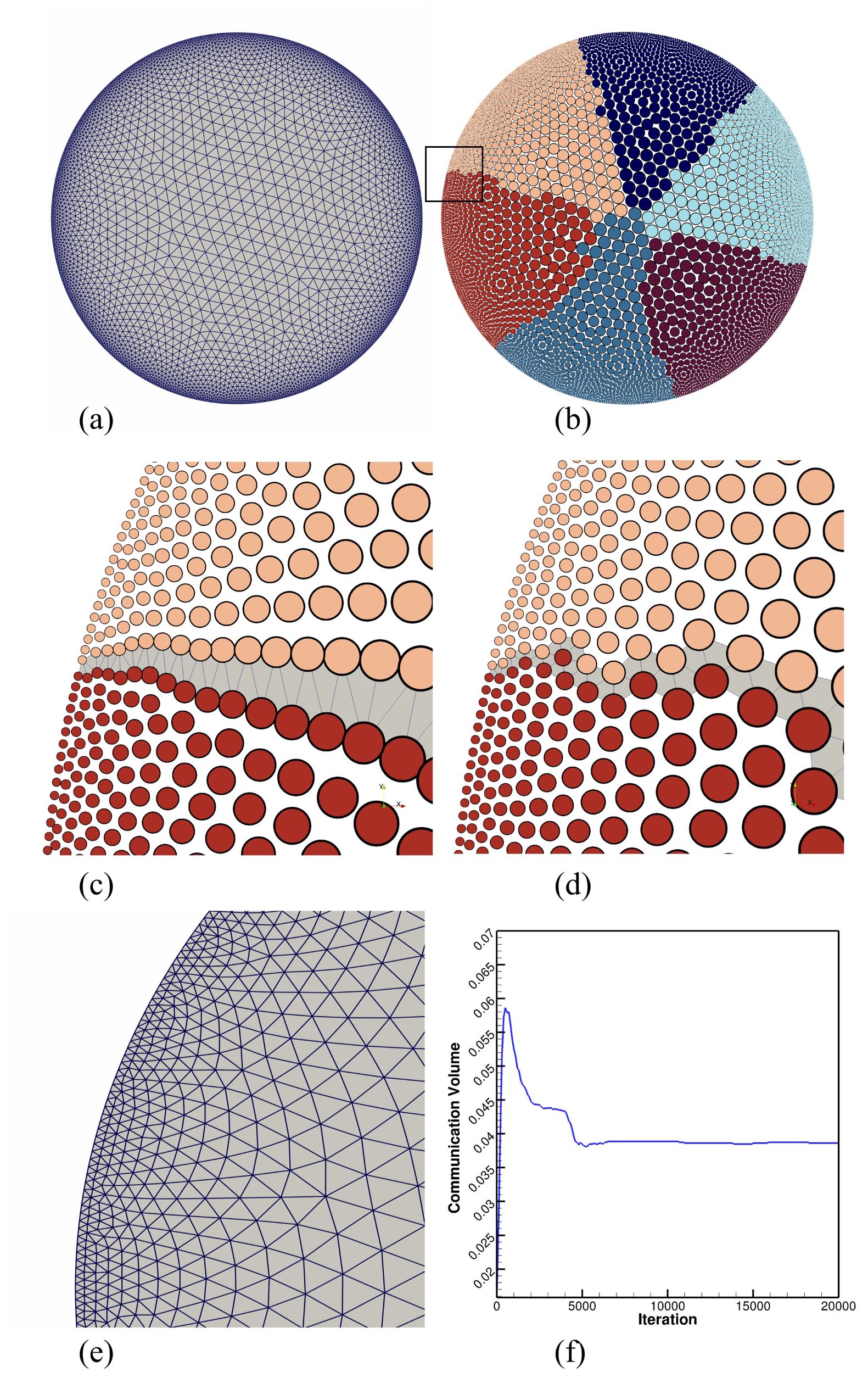}
  \caption{\textit{circle\_6mpi}: (a) Generated mesh after 20000 iterations. (b) Particle distribution after 20000 iterations. Particles are plotted with colors of each sub-domain and radius identical to the target feature-size. Particle distribution (c) before removing surface tension force and (d) after relaxation (zoom-in view of the box region in (b)). (e) Zoomed-in view of the final mesh after 20000 iterations. (f) History of communication volume.}
\label{fig:circle_6mpi_01}
\end{figure}

\begin{figure}[H]
  \centering
    \includegraphics[width=0.8\textwidth]{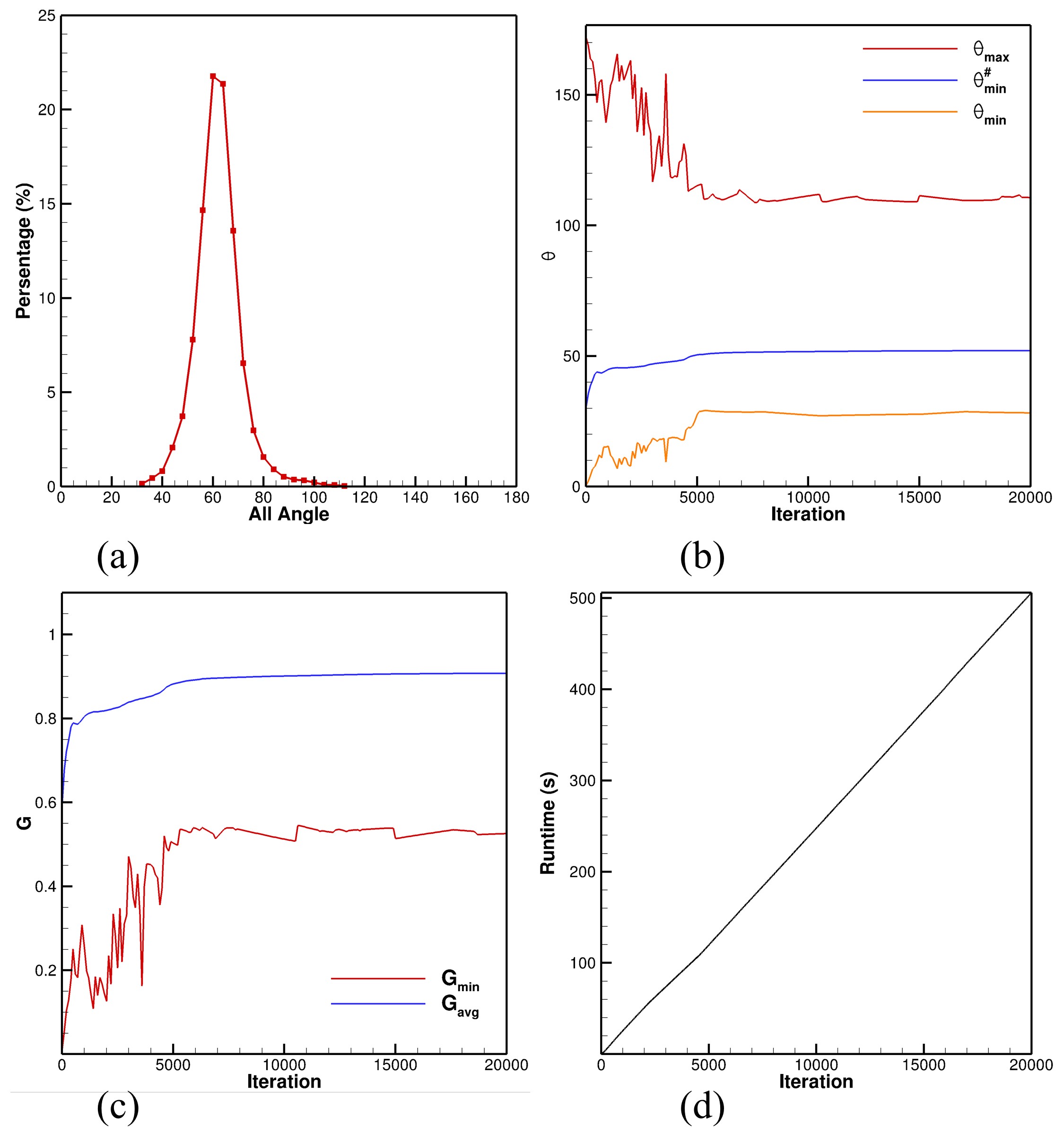}
  \caption{\textit{circle\_6mpi}: (a) Histogram of the angle distribution. (b) Convergence history of $\theta_{max}$, $\theta_{min}$ and $\theta_{min}^{\#}$. (c) Convergence history of $G_{avg}$ and $G_{min}$. (d) History of runtime.}
\label{fig:circle_6mpi_02}
\end{figure}

\begin{figure}[H]
  \centering
    \includegraphics[width=0.8\textwidth]{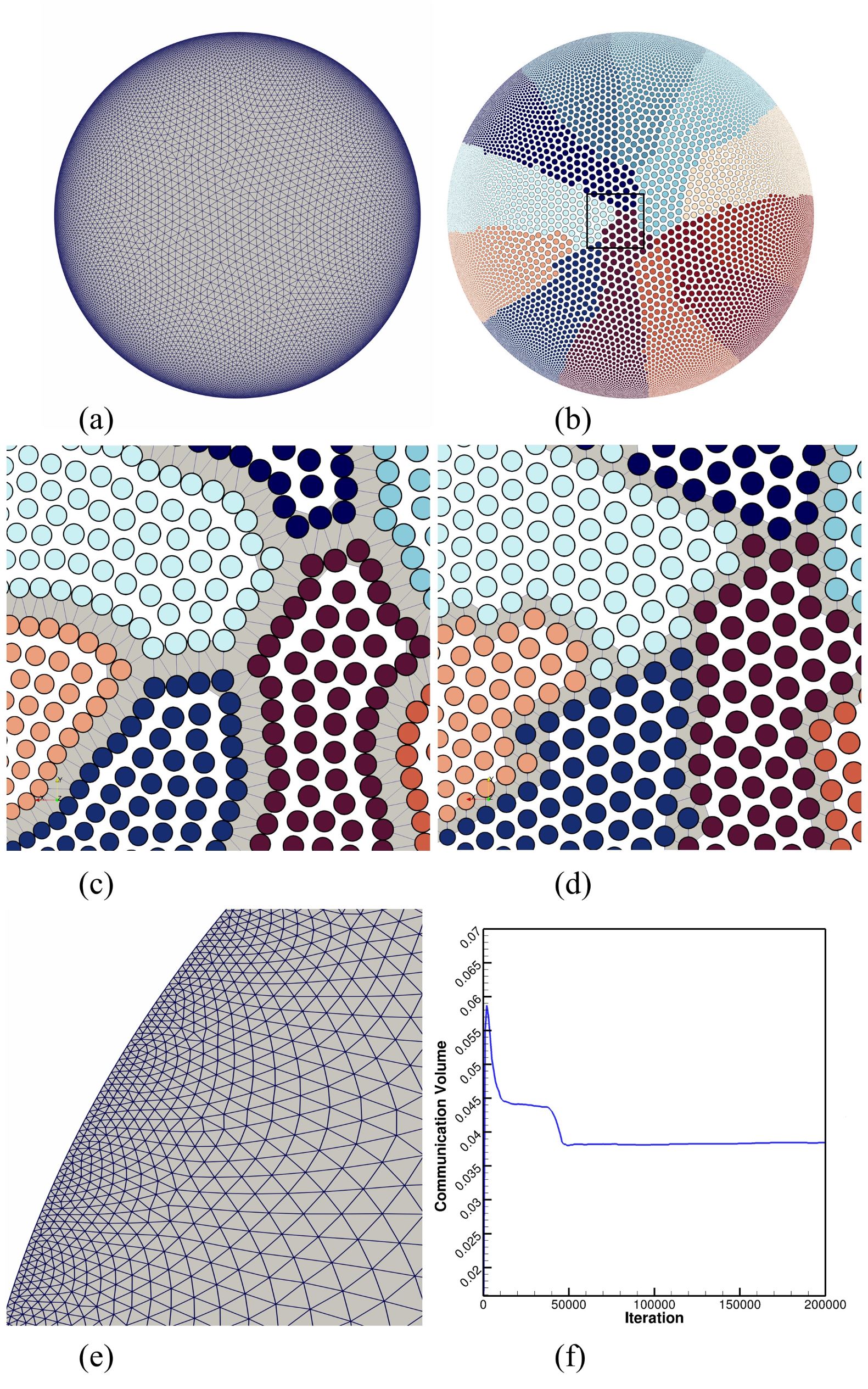}
  \caption{\textit{circle\_12mpi}: (a) Generated mesh after 200000 iterations. (b) Particle distribution after 200000 iterations. Particles are plotted with colors of each sub-domain and radius identical to the target feature-size. Particle distribution (c) before removing surface tension force and (d) after relaxation (zoom-in view of the box region in (b)). (e) Zoomed-in view of the final mesh after 200000 iterations. (f) History of communication volume.}
\label{fig:circle_12mpi_01}
\end{figure}

\begin{figure}[H]
  \centering
    \includegraphics[width=0.8\textwidth]{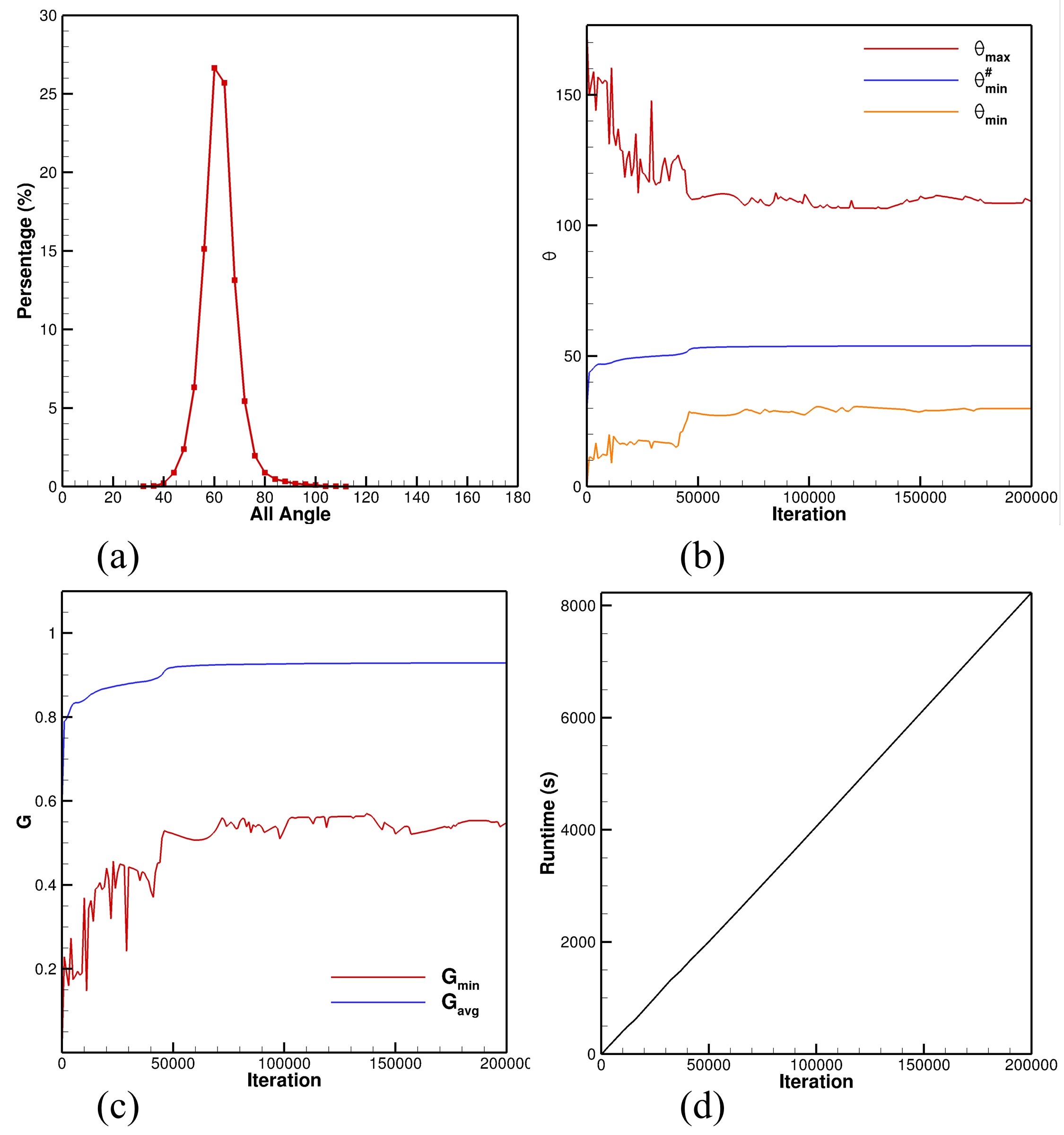}
  \caption{\textit{circle\_12mpi}: (a) Histogram of the angle distribution. (b) Convergence history of $\theta_{max}$, $\theta_{min}$ and $\theta_{min}^{\#}$. (c) Convergence history of $G_{avg}$ and $G_{min}$. (d) History of runtime.}
\label{fig:circle_12mpi_02}
\end{figure}

We first consider a 2D circle with adaptive resolution. The domain is $[0,100]\times[0,100]$, and the circle is defined as 
\begin{equation}
\label{eq:level_set_circle}
	\Gamma=\{(x,y)|43-\sqrt{(x-50)^2+(y-50)^2)}=0\}.
\end{equation}
The target feature-size function is given as $h_t=h_{min}+\frac{tanh(\frac{2.5\phi}{43})}{tanh(2.5)}(h_{max}-h_{min})$, where $h_{max}$ and $h_{min}$ are the maximum and minimum mesh size. Two resolutions are simulated with different number of MPI tasks. For the first case (referred as \textit{circle\_6mpi}), we set $h_{max}=3.125$ and $h_{min}=0.391$, and 6 MPI tasks are launched. For the second case (referred as \textit{circle\_12mpi}), we set $h_{max}=1.95$ and $h_{min}=0.195$, and 12 MPI tasks are launched. The simulation results for \textit{circle\_6mpi} is illustrated in Fig. \ref{fig:circle_6mpi_01} and Fig. \ref{fig:circle_6mpi_02}. Fig. \ref{fig:circle_12mpi_01} and Fig. \ref{fig:circle_12mpi_02} are results of case \textit{circle\_12mpi} respectively. The measurement of mesh quality for both cases are shown in Table \ref{Tab:validation_circle}.

From the simulation results, it can be observed that before the ramping-down of surface tension force, all sub-domains feature convex and connected shape, and a sharp interface is maintained between neighboring sub-domains (see Fig. \ref{fig:circle_6mpi_01} (c) and Fig. \ref{fig:circle_12mpi_01} (c)). After removing the surface-tension force, the sharp-interface condition is gradually relaxed and the mesh vertices near the interface regions are regularized to an isotropic distribution (see Fig. \ref{fig:circle_6mpi_01} (d) and Fig. \ref{fig:circle_12mpi_01} (d)). The final meshes still feature convex shape of sub-domains (see Fig. \ref{fig:circle_6mpi_01} (b) and Fig. \ref{fig:circle_12mpi_01} (b)), and the increase of communication volume after removing the surface tension force is negligible (see Fig. \ref{fig:circle_6mpi_01} (f) and Fig. \ref{fig:circle_12mpi_01} (f)).

High quality meshes are generated for both cases and for both in the regions near geometry boundaries and in the far field (see Fig. \ref{fig:circle_6mpi_01} (a)(e) and Fig. \ref{fig:circle_12mpi_01} (a)(e)). The convergence history of mesh quality and runtime information are also provided in Fig. \ref{fig:circle_6mpi_02} and Fig. \ref{fig:circle_12mpi_02}. Both cases feature proper convergence. Also it can be observed that the overall mesh quality has a rapid increase during the ramping-down of the surface-tension force stage, which starts at approximately 4,000 iterations for \textit{circle\_12mpi} and 40,000 for \textit{circle\_12mpi}). This phenomenon is consistent with the expectation since the repulsive force between sub-domains introduces irregularities at the interface regions.

\begin{table}[h]
\centering
\caption{Mesh quality of the circle case}
\scriptsize
\label{Tab:validation_circle}
\newcommand{\tabincell}[2]{\begin{tabular}{@{}#1@{}}#2\end{tabular}}
\begin{tabular}{>{\centering\arraybackslash}m{1.6cm}
                >{\centering\arraybackslash}m{1cm}
                >{\centering\arraybackslash}m{1cm}
                >{\centering\arraybackslash}m{1cm}
                >{\centering\arraybackslash}m{1cm}
                >{\centering\arraybackslash}m{1cm}
                >{\centering\arraybackslash}m{1cm}
                >{\centering\arraybackslash}m{1cm}
                >{\centering\arraybackslash}m{1cm}}
\hline
 & $G_{avg}$ & $G_{min}$ & $\theta_{max}$ & $\theta_{min}$ & $\theta_{min}^{\#}$ & $\theta_{<30}$ & $N_p$ & $N_{tri}$ \footnotemark \\ \hline
 \textit{circle\_6mpi} & 0.91 &  0.53 &  110.7 & 28.2 & 52.0 & 11 & 4,289 & 7,977 \\
 \textit{circle\_12mpi} & 0.93 &  0.55 &  109.1 & 29.9 & 53.9 & 2 & 13,129 & 25,111 \\
 \hline
\end{tabular}
\end{table}
\footnotetext[1]{$N_p$ denotes the total number of particles and $N_{tri}$ the total number of triangles}

\subsection{Square}
\label{S:validation_square}

\begin{figure}[H]
  \centering
    \includegraphics[width=0.8\textwidth]{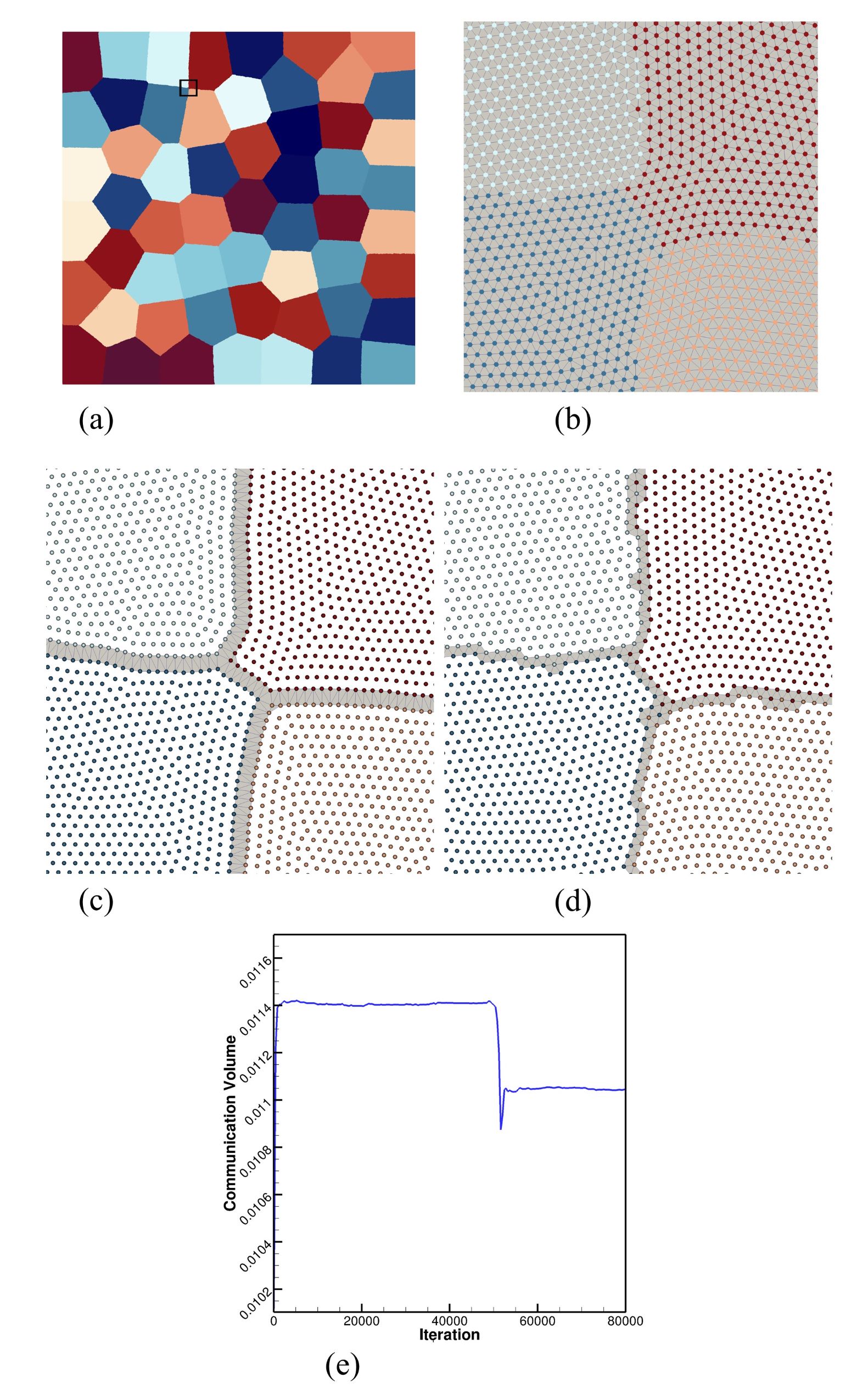}
  \caption{\textit{square\_56mpi}: (a) Generated mesh after 80000 iterations. (b) Zoomed-in view of the box region in (a) after 80000 iterations. Particles are plotted with colors of each sub-domain. Particle distribution (c) before removing surface tension force and (d) after relaxation (zoom-in view of the box region in (a)). (e) History of communication volume.}
\label{fig:square_56mpi_01}
\end{figure}

\begin{figure}[H]
  \centering
    \includegraphics[width=0.8\textwidth]{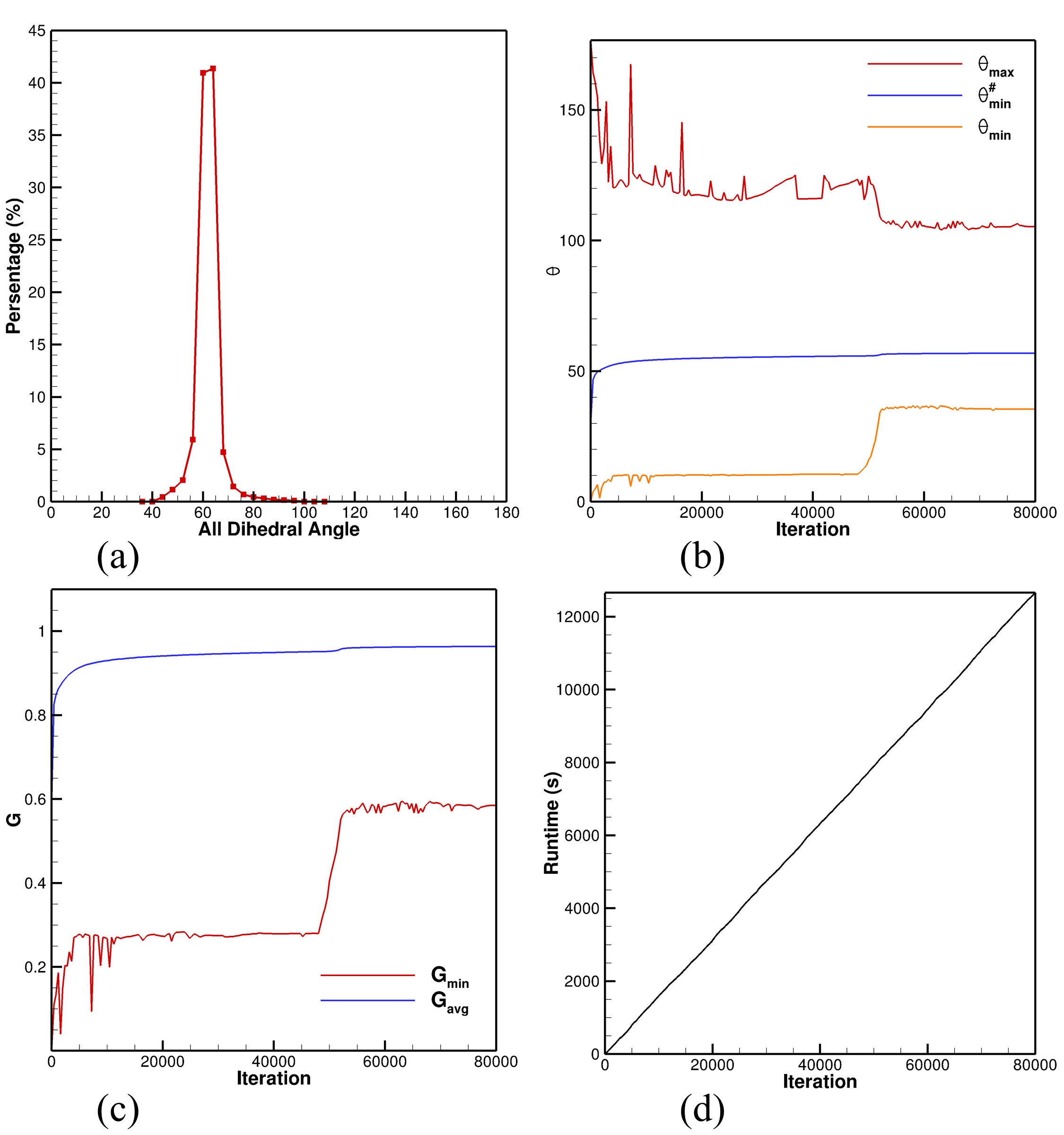}
  \caption{\textit{square\_56mpi}: (a) Histogram of the angle distribution. (b) Convergence history of $\theta_{max}$, $\theta_{min}$ and $\theta_{min}^{\#}$. (c) Convergence history of $G_{avg}$ and $G_{min}$. (d) History of runtime.}
\label{fig:square_56mpi_02}
\end{figure}

Second, we consider a 2D square case with constant resolution using larger scale and larger amount of MPI tasks (referred as \textit{square\_56mpi}). The size of the square is $95\times95$. The total number of mesh vertices is 1,772,894 and 56 MPI tasks are launched. Within each MPI tasks 14 TBB threads are initialized for higher concurrency. The target feature-size is set with a constant value $h_t=0.0714$. For the initial condition, we first initialize 56 Voronoi generators and sample the generators uniformly into the background mesh in positive region. The SPH particles are then randomly sampled inside each Voronoi cell. The simulation results are presented in Fig. \ref{fig:square_56mpi_01} and Fig. \ref{fig:square_56mpi_02} and mesh quality statistics are given in Table \ref{Tab:validation_square}.

The resulting partitioning diagram (see Fig. \ref{fig:square_56mpi_01} (a)(b)) features compact and convex shape of sub-domains, which has high similarities with the partitioning results using the CVP method \cite{fu2017physics} and a graph-based partitioning method in \cite{meyerhenke2009graph}. Similarly with the first case, the sharp-interface condition is preserved before removing the surface tension force (see Fig. \ref{fig:square_56mpi_01} (d)). After optimizing the mesh quality near the interface, small disturbance is introduced at the inter-domain boundaries (see Fig. \ref{fig:square_56mpi_01} (d)), and the amount of communication overhead is limited too (see Fig. \ref{fig:square_56mpi_01} (e)). It is worth mentioning that the optimization of communication with the presence of surface tension force does not always lead to the monotonic decrease of communication volume, which depends on the initial condition particle seeding and the distribution of the target function.

High-quality meshes are obtained from the calculation. Over 80\% of angles range from $55^{\circ}$ to $65^{\circ}$ (see Fig. \ref{fig:square_56mpi_02} (a)). The convergence history of mesh quality measurements presented in Fig. \ref{fig:square_56mpi_02} (b) and (c) demonstrate the good convergence of our method.

\begin{table}[h]
\centering
\caption{Mesh quality of the square case}
\scriptsize
\label{Tab:validation_square}
\newcommand{\tabincell}[2]{\begin{tabular}{@{}#1@{}}#2\end{tabular}}
\begin{tabular}{>{\centering\arraybackslash}m{1.8cm}
                >{\centering\arraybackslash}m{0.8cm}
                >{\centering\arraybackslash}m{0.8cm}
                >{\centering\arraybackslash}m{1cm}
                >{\centering\arraybackslash}m{1cm}
                >{\centering\arraybackslash}m{1cm}
                >{\centering\arraybackslash}m{1cm}
                >{\centering\arraybackslash}m{1cm}
                >{\centering\arraybackslash}m{1cm}}
\hline
 & $G_{avg}$ & $G_{min}$ & $\theta_{max}$ & $\theta_{min}$ & $\theta_{min}^{\#}$ & $\theta_{<30}$ & $N_p$ & $N_{tri}$ \\ \hline
 \textit{square\_56mpi} & 0.96 &  0.58 &  105.3 & 35.5 & 56.9 & 0 & 1,772,894 & 3,541,674 \\
\hline
\end{tabular}
\end{table}

\subsection{Bunny}
\label{S:validation_bunny}

\begin{figure}[H]
  \centering
    \includegraphics[width=0.8\textwidth]{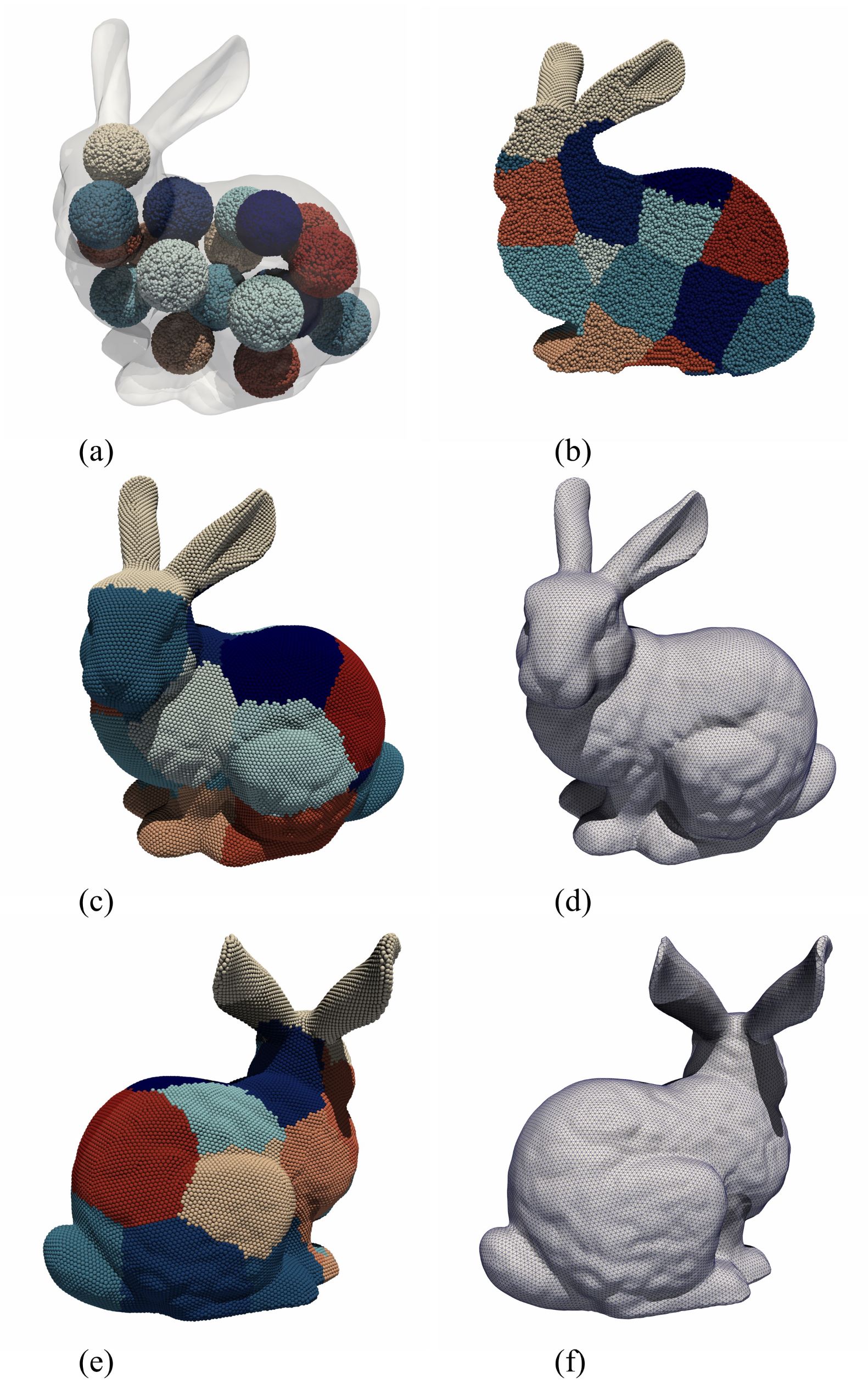}
  \caption{\textit{bunny\_20mpi}: (a) Initial particle distribution. Particles are plotted with sub-domain colors. (b) Cross-section view of the particle distribution after 100,000 iteration. Front view of resulting mesh after 100,000 iteration plotted with (c) particles and (d) surfaces with edges. Back view of resulting mesh after 100,000 iteration plotted with (e) particles and (f) surfaces with edges.}
\label{fig:bunny_20mpi_01}
\end{figure}

The \textit{Stanford Bunny} model \cite{turk1994zippered} is considered. A constant target feature-size of $h_t=1.5$ is defined and the total number of particles is 240,370. 20 MPI tasks are allocated and each contains 14 TBB threads. Similarly with previous case, 20 Voronoi generators are sampled uniformly inside the geometry. Particles are then initialized randomly inside a sphere, whose diameter is the minimum distance to its nearest neighbor. The initial seeding of SPH particles are illustrated in Fig. \ref{fig:bunny_20mpi_01} (a), where particles are rendered with sub-domain colors.

\begin{figure}[H]
  \centering
    \includegraphics[width=0.8\textwidth]{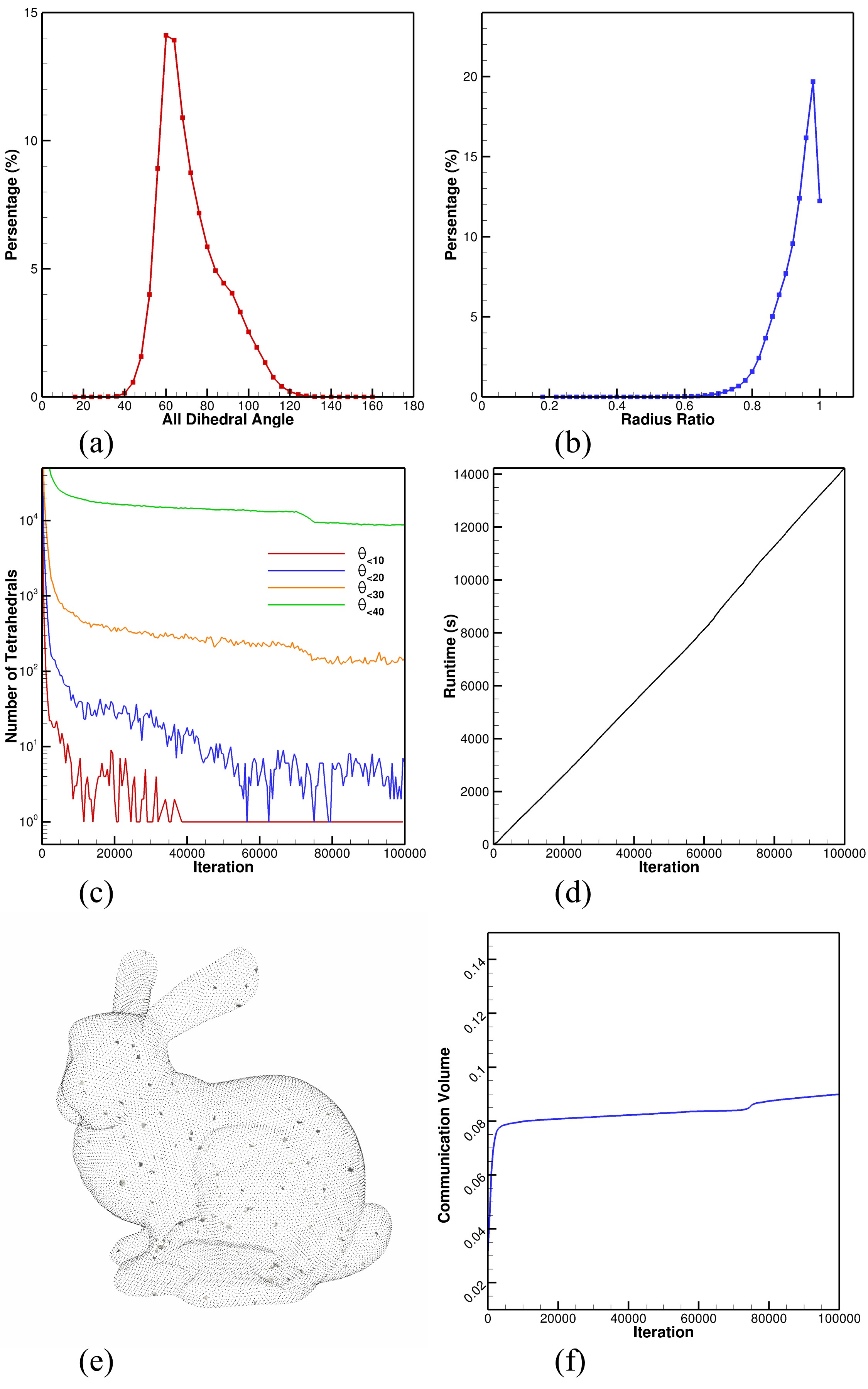}
  \caption{(a) Histogram of the dihedral angle distribution. (b) Histogram of the radius ratio distribution. (c) Convergence history of number of tetrahedra with minimum dihedral angle smaller than $10^{\circ}$, $20^{\circ}$, $30^{\circ}$ and $40^{\circ}$. (d) History of runtime. (e) Tetrahedra with minimum dihedral angle smaller than $30^{\circ}$. (f) History of communication volume.}
\label{fig:bunny_20mpi_02}
\end{figure}

The particle distribution and the generated mesh are illustrated in Fig. \ref{fig:bunny_20mpi_01} (c) to (e) at two different camera positions. A cross-section view is presented as Fig. \ref{fig:bunny_20mpi_01} (b) to show the inter-domain boundaries. All the results demonstrate that the compact and connected sub-domains are maintained, even at the connecting region of the ears. The features and details of the model are well-captured and mesh vertices are distributed homogeneously inside the geometry. The history of communication volume as shown in Fig. \ref{fig:bunny_20mpi_02} (f) exhibits a slight increase after the ramping-down of the surface-tension force (after 75000 iterations), however the overall overhead is approximately 0.4\%.

Good mesh quality is obtained and the statistics are presented in Fig. \ref{fig:bunny_20mpi_02} and Table \ref{Tab:validation_bunny} respectively. The histories of mesh quality demonstrate that the simulation converges properly.

\begin{table}[h]
\centering
\caption{Mesh quality of the Stanford bunny case}
\scriptsize
\label{Tab:validation_bunny}
\newcommand{\tabincell}[2]{\begin{tabular}{@{}#1@{}}#2\end{tabular}}
\begin{tabular}{>{\centering\arraybackslash}m{1cm}
                >{\centering\arraybackslash}m{1.4cm}
                >{\centering\arraybackslash}m{1.4cm}
                >{\centering\arraybackslash}m{0.6cm}
                >{\centering\arraybackslash}m{0.6cm}
                >{\centering\arraybackslash}m{0.6cm}
                >{\centering\arraybackslash}m{0.6cm}
                >{\centering\arraybackslash}m{0.6cm}
                >{\centering\arraybackslash}m{1cm}
                >{\centering\arraybackslash}m{1cm}}
\hline
 & $\theta_{min}/\theta_{max}$ & $\gamma_{min}/\gamma_{avg}$ & $\theta_{min}^{\#}$ & $\theta_{<10}$ & $\theta_{<20}$ & $\theta_{<30}$ & $\theta_{<40}$ & $N_p$ & $N_{tet}$ \\ \hline
 \textit{bunny} \textit{\_20mpi} & 15.3/154.8 & 0.24/0.92 & 56.6 & 0 & 2 & 127 & 8636 & 240,370 & 1,366,196 \\
\hline
\end{tabular}
\end{table}

\subsection{Cube}
\label{S:validation_cube}

\begin{figure}[H]
  \centering
    \includegraphics[width=0.8\textwidth]{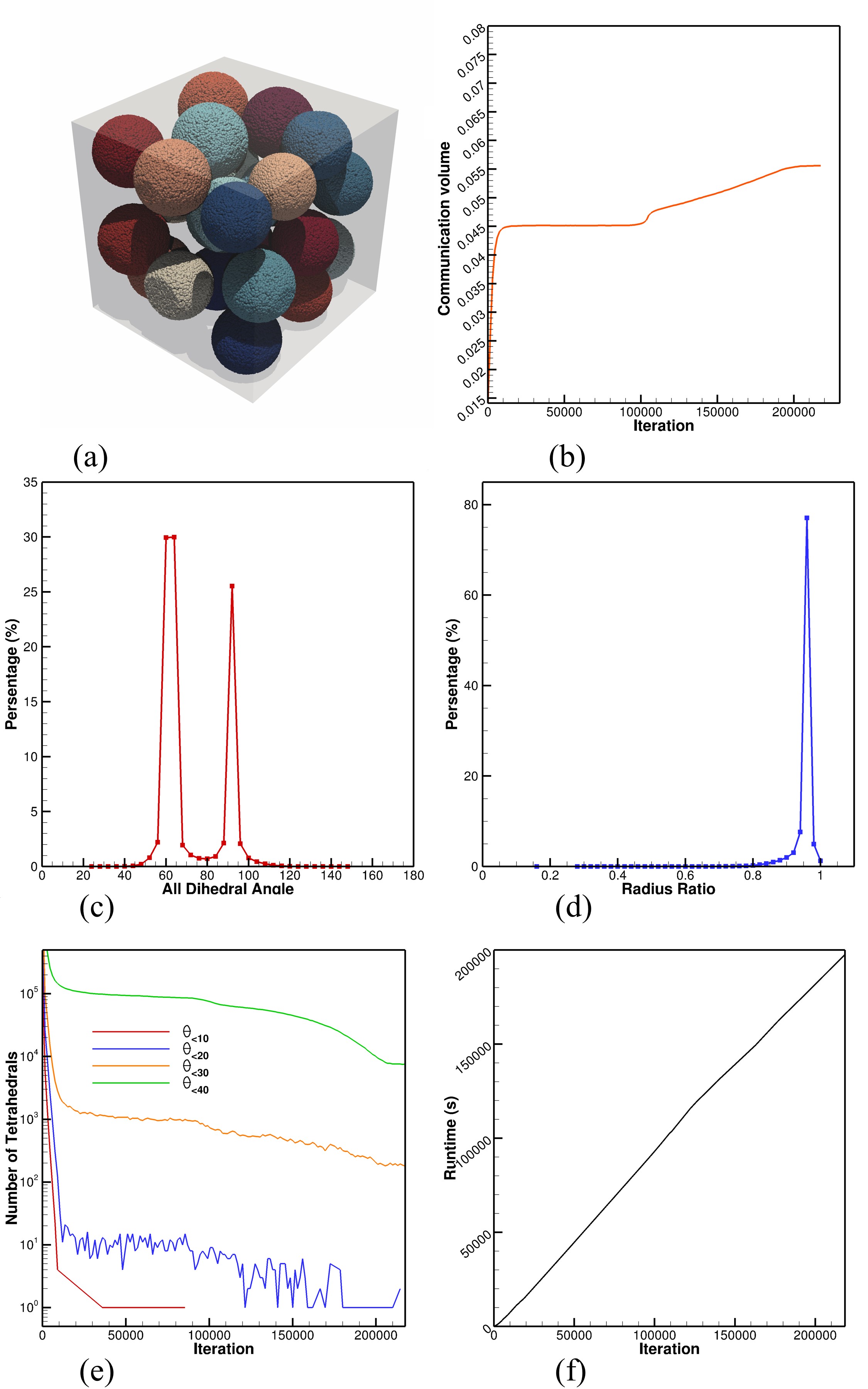}
  \caption{(a) Initial seeding of particles. (b) History of communication volume. (c) Histogram of the dihedral angle distribution. (d) Histogram of the radius ratio distribution. (e) Convergence history of number of tetrahedra with minimum dihedral angle smaller than $10^{\circ}$, $20^{\circ}$, $30^{\circ}$ and $40^{\circ}$. (f) History of runtime.}
\label{fig:cube_24mpi_02}
\end{figure}

We consider a cube of size $[95\times95\time95]$. This is a simple geometry but contains both singularities and feature curves. A constant target feature-size of $h_t=0.78$ is defined and the total number of particles is 1,888,113. 24 MPI tasks are allocated and each contains 14 TBB threads. Similarly with the bunny case, 24 Voronoi generators are sampled uniformly inside the geometry. The initial particle seeding is shown in Fig. \ref{fig:cube_24mpi_02} (a).

\begin{figure}[H]
  \centering
    \includegraphics[width=0.8\textwidth]{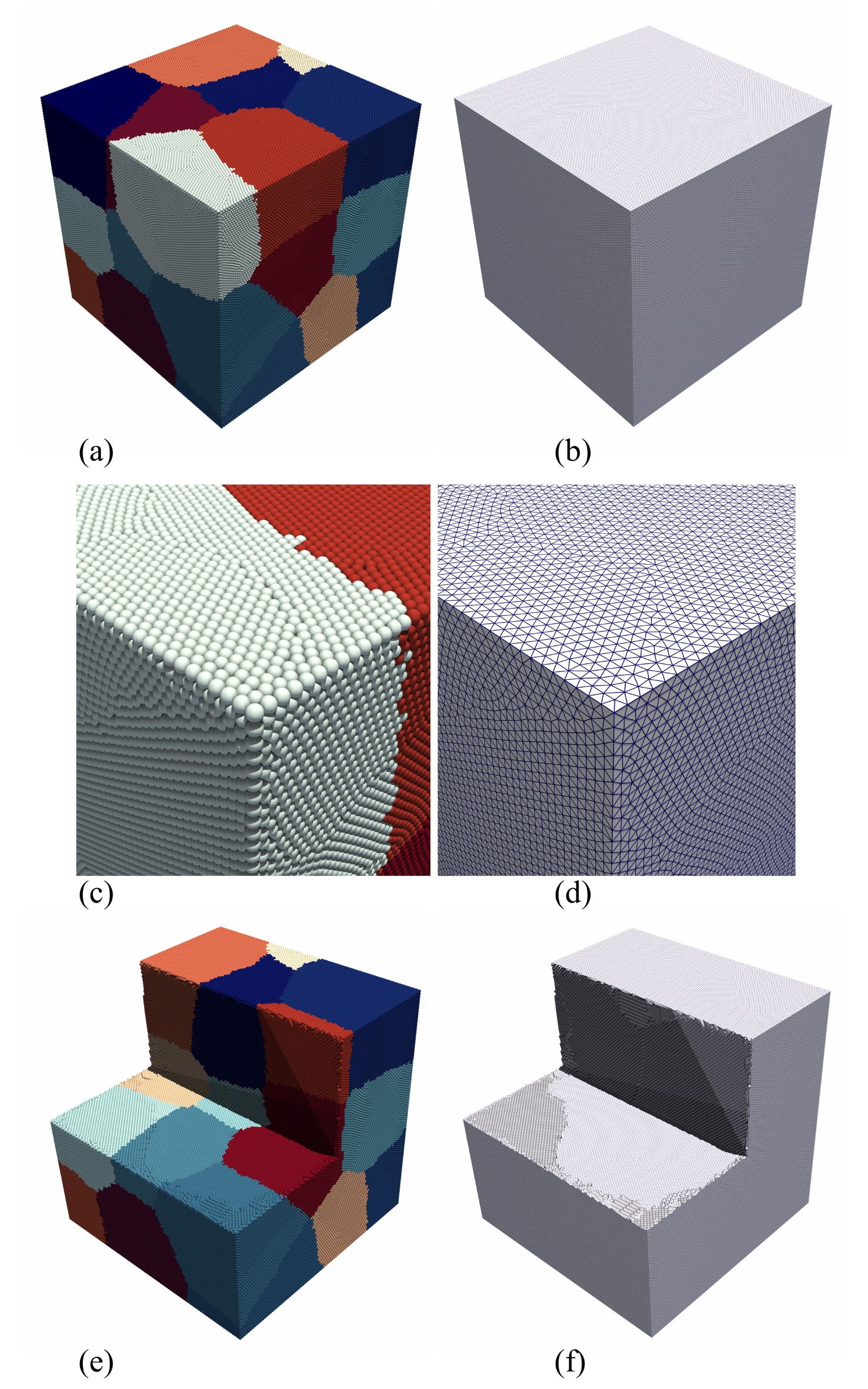}
  \caption{Simulation result after 217500 iterations plotted with (a) particles and (b) surfaces with edges. (c) Zoomed-in view of (a). (d) Zoomed-in view of (b). Simulation result with clipping after 217500 iterations plotted with (e) particles and (f) surfaces with edges. Particles are rendered by sub-domain colors.}
\label{fig:cube_24mpi_01}
\end{figure}

The simulation result after 217500 iterations is illustrated in Fig. \ref{fig:cube_24mpi_01} presented by particles rendered with sub-domain colors (see Fig. \ref{fig:cube_24mpi_01} (a)(c)(e)) and tetrahedra (see Fig. \ref{fig:cube_24mpi_01} (b)(d)(f)). Again overall compact and convex sub-domains are maintained with small disturbance at the interface (see Fig. \ref{fig:cube_24mpi_01} (c) for a zoom-in view and Fig. \ref{fig:cube_24mpi_01} (e) for a clipped view). The overhead of communication volume after relaxation of surface tension force is approximately 1\%.

Regarding to mesh quality, statistics and history curves are presented in Table \ref{Tab:validation_cube} and Fig. \ref{fig:cube_24mpi_02}. According to the Dihedral angle histogram in Fig. \ref{fig:cube_24mpi_02} (c), most of the angles concentrate at approximately $60^{\circ}$ ($\sim60\%$) and $90^{\circ}$ ($\sim25.5\%$), which is highly close to the dihedral angles of a Body-Centered-Cubic (BCC) tetrahedron and features minimum mean square error \cite{barnes1983optimal}. From the zoom-in view (see Fig. \ref{fig:cube_24mpi_01} (d)) and clipped view of the generated mesh (see Fig. \ref{fig:cube_24mpi_01} (f)), the BCC distribution of mesh vertices can be observed. For the radius ratio diagram (see Fig. \ref{fig:cube_24mpi_02} (d)), about 83\% of all tetrahedra falls into the range between 0.94 to 1.13.

\begin{table}[h]
\centering
\caption{Mesh quality of the cube case}
\scriptsize
\label{Tab:validation_cube}
\newcommand{\tabincell}[2]{\begin{tabular}{@{}#1@{}}#2\end{tabular}}
\begin{tabular}{>{\centering\arraybackslash}m{1cm}
                >{\centering\arraybackslash}m{1.4cm}
                >{\centering\arraybackslash}m{1.4cm}
                >{\centering\arraybackslash}m{0.6cm}
                >{\centering\arraybackslash}m{0.6cm}
                >{\centering\arraybackslash}m{0.6cm}
                >{\centering\arraybackslash}m{0.6cm}
                >{\centering\arraybackslash}m{0.6cm}
                >{\centering\arraybackslash}m{1cm}
                >{\centering\arraybackslash}m{1cm}}
\hline
 & $\theta_{min}/\theta_{max}$ & $\gamma_{min}/\gamma_{avg}$ & $\theta_{min}^{\#}$ & $\theta_{<10}$ & $\theta_{<20}$ & $\theta_{<30}$ & $\theta_{<40}$ & $N_p$ & $N_{tet}$ \\ \hline
 \textit{cube} \textit{\_24mpi} & 21.5/147.3 & 0.27/0.94 & 58.8 & 0 & 0 & 187 & 7796 & 1,888,113 & 11,127,061 \\
\hline
\end{tabular}
\end{table}

\subsection{Spur gear}
\label{S:validation_gear}

\begin{figure}[H]
  \centering
    \includegraphics[width=0.8\textwidth]{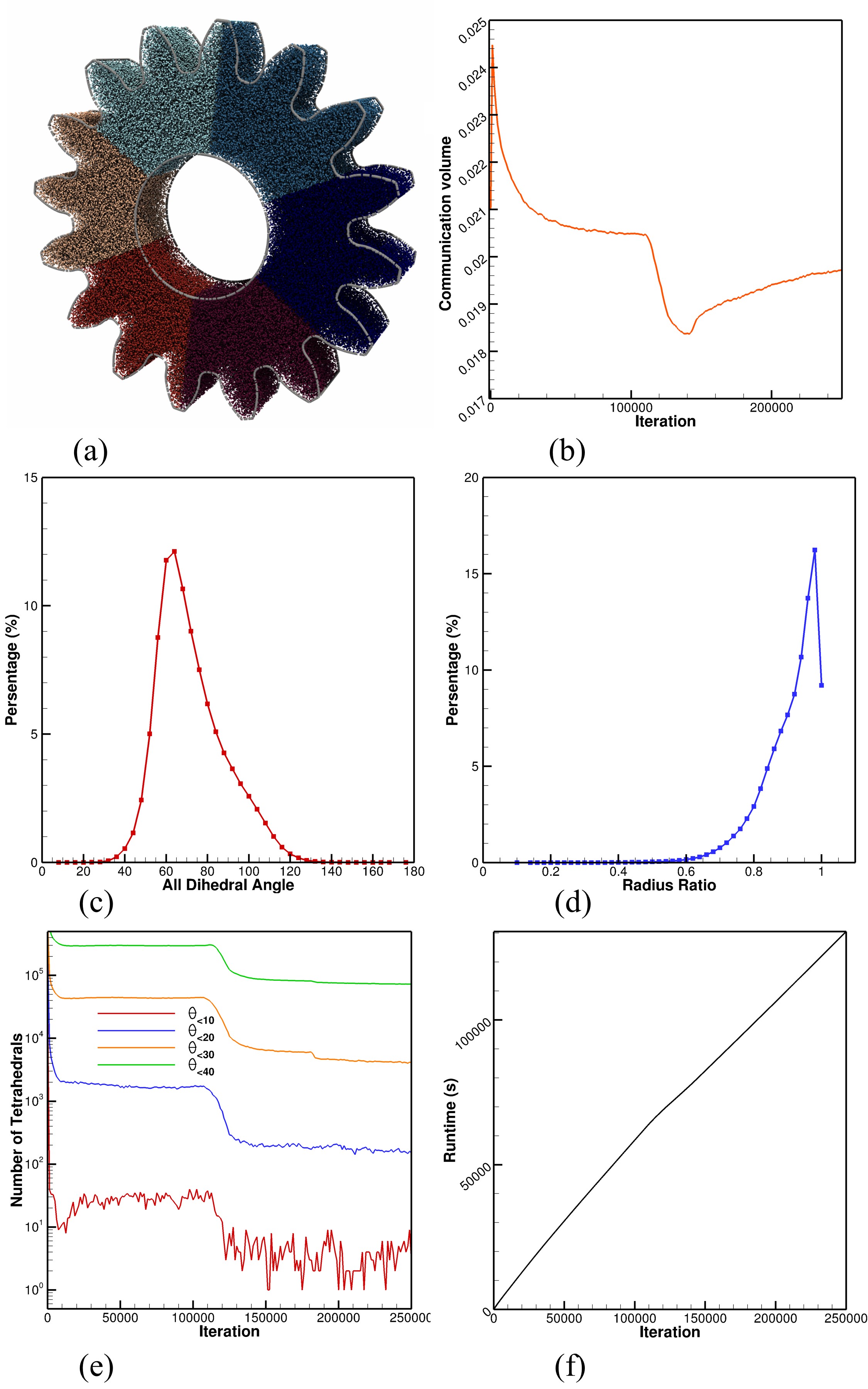}
  \caption{(a) Initial seeding of particles. (b) History of communication volume. (c) Histogram of the dihedral angle distribution. (d) Histogram of the radius ratio distribution. (e) Convergence history of number of tetrahedra with minimum dihedral angle smaller than $10^{\circ}$, $20^{\circ}$, $30^{\circ}$ and $40^{\circ}$. (f) History of runtime.}
\label{fig:FZG_gear_6mpi_02}
\end{figure}

Lastly we consider a realistic geometry of spur gear developed for gear lubrication tests \cite{hoehn2008test}. Multiple singularities and sharp edges are presented in the model. The size of computational domain is $[86.4\times17.6\times86.4]$. The minimum and maximum target feature-size is 0.4 and 1.6 respectively. The target density field is calculated considering singularities, feature curve, curvature and distance to the geometry surface similar with \cite{FU2019396}. The total number of particles calculated is 1,888,113. 6 MPI tasks are allocated and each contains 14 TBB threads. 6 Voronoi generators are uniformly distributed in the geometry and particles are initially randomly sampled within each Voronoi cell. The initial condition is shown in Fig. \ref{fig:FZG_gear_6mpi_02} (a).

\begin{figure}[H]
  \centering
    \includegraphics[width=0.8\textwidth]{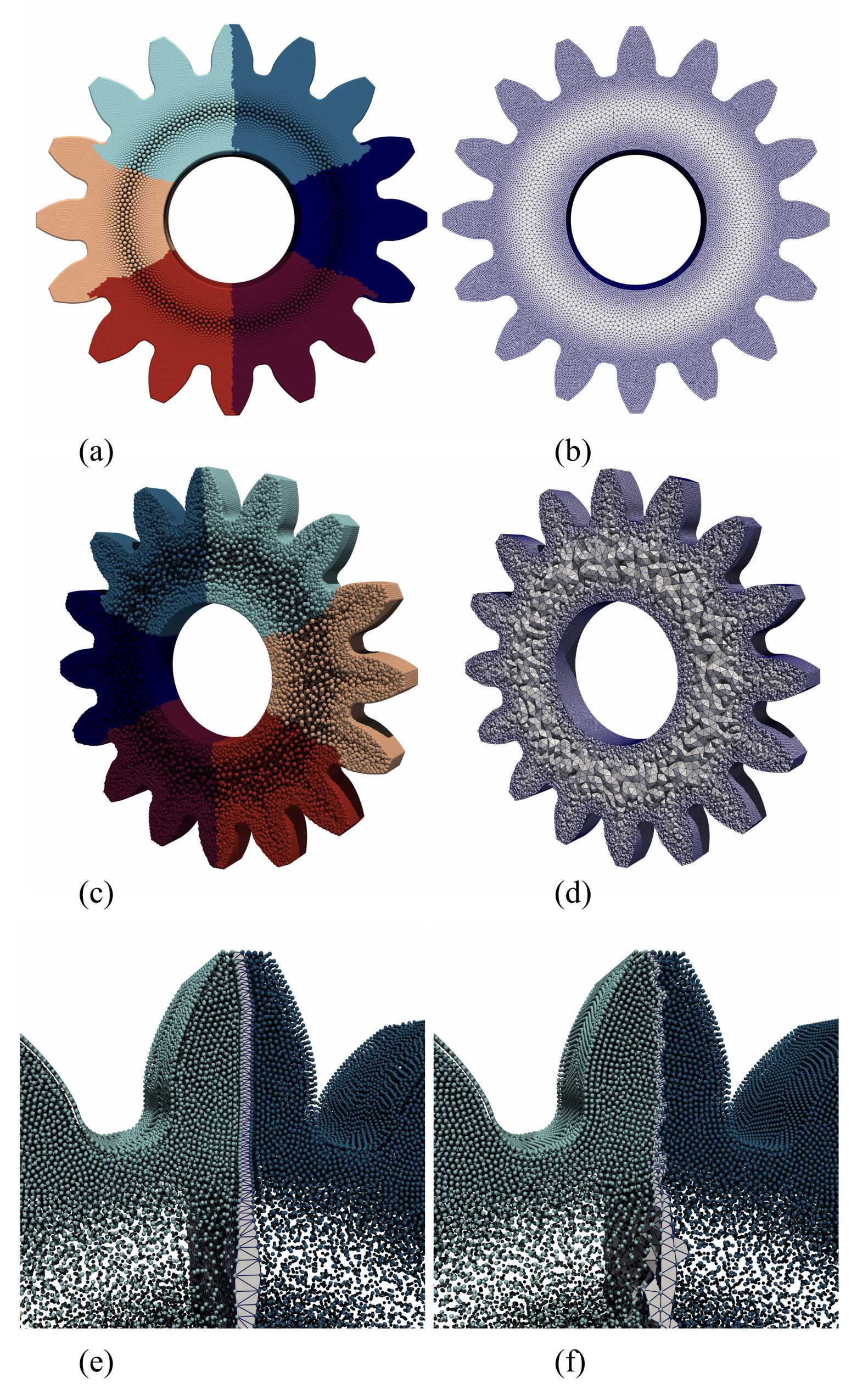}
  \caption{Simulation result after 250000 iterations plotted with (a) particles and (b) surfaces with edges. (c) Clipped view of (a). (d) Clipped view of (b). Particle distribution (c) before removing surface tension force and (d) after fully relaxed. Particles are rendered by sub-domain colors.}
\label{fig:FZG_gear_6mpi_01}
\end{figure}

The generated results after 250000 iterations are illustrated in Fig. \ref{fig:FZG_gear_6mpi_01} (a) and (b) with particle and mesh representation, and clipped views (see Fig. \ref{fig:FZG_gear_6mpi_01} (c)(d)) are presented too. The resulting partitioning sub-domains feature convex and compact shape. The sharp-interface condition is maintained before the relaxation of surface-tension force (see Fig. \ref{fig:FZG_gear_6mpi_01} (e)). While some disturbances are observed at the geometry corners after the system is fully relaxed (see Fig. \ref{fig:FZG_gear_6mpi_01} (a) and {f}), the communication overhead is around 0.15\% (see \ref{fig:FZG_gear_6mpi_02} (b)).

The histories of $\theta_{<10}$, $\theta_{<20}$, $\theta_{<30}$ and $\theta_{<40}$ (\ref{fig:FZG_gear_6mpi_02} (e)) show good convergence of the simulation. The number of slivers is ignorable, i.e. 1 tetrahedron has dihedral angle smaller than $10^{\circ}$, comparing to total number of elements generated (1,847,416). The histograms of dihedral angle (\ref{fig:FZG_gear_6mpi_02} (c)) and radius ratio (\ref{fig:FZG_gear_6mpi_02} (d)) exhibit good mesh quality too.

\begin{table}[h]
\centering
\caption{Mesh quality of the spur gear case}
\scriptsize
\label{Tab:validation_FZG_gear}
\newcommand{\tabincell}[2]{\begin{tabular}{@{}#1@{}}#2\end{tabular}}
\begin{tabular}{>{\centering\arraybackslash}m{1.5cm}
                >{\centering\arraybackslash}m{1.4cm}
                >{\centering\arraybackslash}m{1.4cm}
                >{\centering\arraybackslash}m{0.6cm}
                >{\centering\arraybackslash}m{0.5cm}
                >{\centering\arraybackslash}m{0.5cm}
                >{\centering\arraybackslash}m{0.5cm}
                >{\centering\arraybackslash}m{0.5cm}
                >{\centering\arraybackslash}m{1cm}
                >{\centering\arraybackslash}m{1cm}}
\hline
 & $\theta_{min}/\theta_{max}$ & $\gamma_{min}/\gamma_{avg}$ & $\theta_{min}^{\#}$ & $\theta_{<10}$ & $\theta_{<20}$ & $\theta_{<30}$ & $\theta_{<40}$ & $N_p$ & $N_{tet}$ \\ \hline
 \textit{Spur\_gear} \textit{\_6mpi} & 6.84/168.7 & 0.14/0.90 & 54.8 & 1 & 162 & 4244 & 73411 & 358,836 & 1,847,416 \\
\hline
\end{tabular}
\end{table} 

\section{Conclusions}
\label{S:conclusion}

In this paper, we have developed a consistent parallel mesh generation method with a multi-phase SPH formulation and particle relaxation strategy. The objectives of partitioning the domain, optimizing communication volume and improving mesh quality are achieved consistently by solving the same set of physics-motivated governing equations. The main contributions of the paper are:

\begin{description}
	\item[(1)] A unified target density function is defined to characterize the targets of both the domain decomposition and the mesh generation. The target density function can be any smooth scalar field considering various geometrical features and user-defined inputs. By utilizing a background Cartesian mesh and level-set function, the total number of mesh vertices and target mass for each sub-domain can be determined a priori;
	\item[(2)] A parallelization strategy is developed and a set of physics motivated governing equations are proposed to achieve all underlying targets consistently. A surface tension model is introduced to the previous particle-based mesh generator \cite{FU2019396} to handle the additional target of optimizing the communication volume in a parallel environment. During the domain decomposition stage, the mesh quality is improved simultaneously in the interior region of each sub-domain. Once a steady state is achieved, the mesh quality near the interface region is optimized by gradually alleviating the surface tension force.
	\item[(3)] A multi-phase SPH formulation is utilized to solve the governing equations. The previously-developed mesh generator \cite{FU2019396} is extended to higher dimensions and parallelized with both MPI and TBB technique. Numerical results demonstrate that the resulting sub-domains feature compact and regularized shape, and high-quality mesh is generated simultaneously. The communication overhead caused by the optimization of mesh quality near the interface is limited even in cases with complex geometry and large spacial adaptivity;
	\item[(4)] With the proposed parallel mesh generation method, high quality triangle/tetrahedron mesh can be generated without the need of constructing Delaunay Triangulation/Tetrahedralization explicitly. Since only local information are required during the simulation and the same set of governing equations are solved for all the particles, the proposed method features high consistency and code reusability. Benefiting from the scalable parallel environment designed previously in \cite{ji2018new}, the mesh generation procedure is able to exploit both fine-grained and coarse-grained parallelization.
\end{description}

Given all the above-mentioned advantages, the current particle-based method is still considerably more expensive than the state-of-the-art Delaunay-based methods. In the future, we are looking forward to extend the proposed algorithm to GPU-based architectures to achieve higher concurrency and performance, since most SPH methods have experienced one to two orders of magnitude speedups when extended to GPUs. Moreover, more studies on initial particle seeding strategies will be carried out and possible coupling with existing mesh-generation methods will be investigated in the near future.


\label{}



\section*{Acknowledgments}

Zhe Ji is partially supported by China Scholarship Council (No. 201506290038). Xiangyu Hu acknowledges funding Deutsche Forschungsgemeinschaft (HU1527/10-1 and HU1527/12-1). The computational resources are provided by Leibniz-Rechenzentrum der Bayerische Akademie der Wissenschaften, Munchen (LRZ).



  \bibliographystyle{elsarticle-num}
  \scriptsize
  \setlength{\bibsep}{0.5ex}

\bibliography{bib}

\end{document}